\documentclass[pdf,aoas,preprint]{imsart}   

\usepackage{amsthm,amsmath,natbib, verbatim}
\usepackage{epsfig,graphicx,amssymb,times}

\RequirePackage[colorlinks,citecolor=blue,urlcolor=blue]{hyperref}
\RequirePackage{hypernat}

\startlocaldefs
\newtheorem{theorem}{Theorem}[section]

\newtheorem{proposition}[theorem]{Proposition}

\endlocaldefs

\begin{document}

\begin{frontmatter}
\title{Inference and Characterization of Multi-Attribute Networks with Application to Computational Biology\protect\thanksref{T1}}
\runtitle{Multi-Attribute Networks}

\begin{aug}
\author{\fnms{Natallia} \snm{Katenka}\ead[label=e1]{nkatenka@bu.edu}} and
\author{\fnms{Eric D.} \snm{Kolaczyk}\ead[label=e2]{kolaczyk@math.bu.edu}}
\thankstext{T1}{This work was supported in part by NIH award GM078987 and by ONR award N000140910654.}
%\thankstext{t1}{Corresponding author }
%\thankstext{t2}{First supporter of the project}
%\thankstext{t3}{Second supporter of the project}
\runauthor{Katenka and Kolaczyk}
\affiliation{Boston University}  %   \thanksmark{m1}}
\address{Department of Mathematics and Statistics\\
111 Cummington Street, Boston, MA 02215\\
\printead{e1}\\
\phantom{E-mail:nkatenka@bu.edu}}
\address{Department of Mathematics and Statistics\\
111 Cummington Street, Boston, MA 02215\\
\printead{e2}\\
\phantom{E-mail:kolaczyk@math.bu.edu}}

\end{aug}

\begin{abstract}

 Our work is motivated by and illustrated with application of association networks in computational biology, specifically in the context of gene/protein regulatory networks. Association networks represent systems of interacting elements, where a link between two different elements indicates a sufficient level of similarity between element attributes. While in reality relational ties between elements can be expected to be based on similarity across multiple attributes, the vast majority of work to date on association networks involves ties defined with respect to only a single attribute.  We propose an approach for the inference of multi-attribute association networks from measurements on continuous attribute variables, using canonical correlation and a hypothesis-testing strategy.  Within this context, we then study the impact of partial information on multi-attribute network inference and characterization, when only a subset of attributes is available.  We consider in detail the case of two attributes, wherein we examine through a combination of analytical and numerical techniques the implications of the choice and number of node attributes on the ability to detect network links and, more generally, to estimate higher-level network summary statistics, such as node degree, clustering coefficients, and measures of centrality. Illustration and applications throughout the paper are developed using gene and protein expression measurements on human cancer cell lines from the NCI-60 database.

\end{abstract}

\end{frontmatter}

\section{Introduction}
\label{sec:introduction}

  Networks have been used for mathematical representation of systems of interacting elements in the context of a wide range of technological, biological, and social applications. Statistical analysis of network data has become particularly popular in the past decade.  See~\cite{Kolaczyk}, for example, for a comprehensive overview of the main classes of methods for statistical inference on networks, as well as~\cite{Goldenberg},- for a shorter review. Although the results presented in this paper are applicable to various network applications, our current work has been motivated by and will be illustrated within the context of gene/protein regulatory networks. Regulatory interactions among genes/proteins are pivotal to the function of living organisms, and understanding regulatory networks can help to characterize  biological processes in general, and also to diagnose different diseases and develop new cures. 

The standard representation for a network is a graph that consists of a collection of nodes (e.g., genes, proteins, social actors, computers) and links that indicate some notion of node interaction (e.g., co-regulation, interaction, friendship, communication). Additionally, nodes or links, or both, can be accompanied by a single or a set of multiple attributes or characteristics. One of the fundamental problems in the area, common across different applications, is that of inferring the underlying network topology. Examples arise in the context of gene/protein regulatory networks, computer networks, sensor networks, social networks, and more. For example, based on observed flow data between different computers, a reasonable communication network can be approximated (e.g., \cite{ErikssonEtAl:imc2007}); based on obtained geographical positions, a randomly deployed wireless sensor network can be reconstructed (e.g., \cite{Pal_2010}); or based on data gathered from individuals about their personal interaction, preference and/or attitudes, a network of social relations can be produced (e.g,  \cite{Sampson_1969}).            

There are a number of variations on the problem of network topology inference.  See ~\citet[Chap. 7]{Kolaczyk}, for example, for an overview.  In this paper, we focus on inference of {\em association networks}, where a link between two different nodes is said to exist when a sufficient level of association is present  between a certain set of node characteristics (attributes). A link between two nodes in an association network may indicate a certain level of interaction, dependence, or similarity, depending on how `association' is quantified.   While in reality the actual  relational ties between elements typically are based on association across multiple attributes, the vast majority of work to date on association networks involves ties defined with respect to only a single attribute. Here we are interested in recovering the structure of an association network where multiple attributes are observed for each node. 

Analysis of multiple attributes at their corresponding network links  has received comparatively little attention in the literature. In the early 1980s log-linear models were adapted by \cite{Fienberg_1985} for the analysis of social interaction networks among 18 monks in a cloister and the analysis of a corporate interlock network of the 25 largest organizations in Minneapolis/St.Paul; much later, canonical correlation analysis was applied  by \cite{Carol_2006} to two multiplex networks that described interdependence and cooperative alliances between 317 banks. Other examples  include work predicting friendships, the participation of actors in events,- and semantic relationships such as 'advisor-of' based on web page links and content (see \cite{Goldenberg} for more a detailed review). More recently, \cite{Chang_2010} focused on multiple attributes of document networks and developed a hierarchical model of both network structure and node attributes. Using repeated interactions between senders and receivers  tabulated over time \cite{Perry_2011} modeled message sending behavior in a corporate e-mail network. Although these studies are focused on the analysis of networks equipped with multiple node attributes, they differ in a critical manner from our work in that they assume observed network topologies, rather than -- as here -- focusing specifically on the problem of inferring the network from the node attributes. 

 The importance of this distinction is particularly evident within the context of computational biology and our motivating application therein. In particular, current and anticipated `Omic' technologies (e.g., genomics, transcriptomics, proteomics and metabolomics) can profile cells at different biological levels, including but not limited to gene, protein, metabolic, and epigenetic levels. While computational  analyses (e.g., differential expression, clustering, network, etc.) based on individual types of profiles have no doubt proven to be useful, analyses based on multiple types of molecular profiles combined on the same set of biological samples can be synergistic. See, for example, \cite{Lee_2004, Myers_2005, Shankavaram2007, Waaijenborg_2008, Naylor_2010}. The work in~\cite{Lee_2004} is perhaps closest in spirit to ours, in that multiple networks initially inferred from diverse single functional genomics data are integrated to form a single network, using a log-likelihood scoring scheme. 

To the best of our knowledge, there has been no work on direct inference of multi-attribute networks with particular attention to specifically understanding (a) how different node attributes contribute to the strength of a link between different nodes and (b) the impact of  having available only a subset of attributes, both on the inference of network topology and the interpretation of high-level network characteristics. In the research we report here we address these issues by answering the following questions: how to aggregate observed multiple continuous attribute variables into a single measure of the total similarity; how to assess the contribution of each node attribute to this similarity measure; what the implications of the choice and the number of node attributes are on high-level network characteristics, such as node degree, clustering coefficient, and betweenness centrality; and, finally, how to extract and interpret information obtained from a network inferred from multiple node attributes.  

Specifically, to aggregate multiple attributes into a measure of a total similarity between a pair of nodes, we propose to quantify the strength of the link between different nodes with canonical correlation, originally introduced by \cite{Hotelling1936}. Within this context, we then examine both analytically and numerically the impact of partial information on the ability to detect a link between a pair of nodes. To assess the importance of individual node attributes, we use a notion of canonical weights. We explore the impact of the attribute selection on higher-level network summary statistics in the context of gene/protein regulatory networks in human cancer cells. Finally, based on the association network inferred from combined profiles of genes and proteins, we propose a simple heuristic for link and node classification  that allows to make reasonable interpretation of the connection between attributes and classified nodes. We validate the proposed heuristic by determining the significant enrichments of known genomic entities among acquired classes of nodes.

  The rest of the paper is structured as follows. In Section \ref{sec:motivation} we introduce the motivating application of our study and describe related work in the area. In Section \ref{sec:methods} we provide a general formulation of the problem, state the main assumptions, and introduce the mathematical notion of canonical correlation in terms of network inference. In Section~\ref{sec:link_declaration} we describe a method of network inference based on hypothesis testing and we explore the effect of different parameters on the power of link detection. In Section \ref{sec:application} we study potential implications of node attribute selection on network summary statistics in the context of gene/protein regulatory networks. We conclude the paper with the discussion and final remarks in Section \ref{sec:remarks}.

\section{Motivating Application}
\label{sec:motivation}

 In the application herein, we explore the use of multi-attribute association network analysis for combining measurements on gene and protein expression levels in order to recover networks of gene/protein interactions effectively.  

 We choose to analyze data from the well-known NCI-60 database, which contains different molecular profiles on a panel of 60 diverse human cancer cell lines\footnote{Dataset available at http://discover.nci.nih.gov/}. Specifically, we examine protein profiles (i.e., normalized reverse-phase lysate arrays (RPLA) for 92 antibodies) and gene profiles (i.e., normalized RNA microarray intensities from Human Genome U95 Affymetrix chip-set for $> 9000$ genes). Traditionally, it has been significantly more difficult to obtain protein-level expression measurements than gene-level expression measurements, although the former typically have been considered to be more accurate and informative than the latter. Accordingly, our analysis will be restricted to a common subset of $91$ genes/proteins for which both
types of biological measurements are available to us.  Each gene/protein is represented by its Entrez ID (a unique identifier common for a protein and a corresponding gene that encodes this protein)  and has a pair of attributes: protein profile and gene expression across the same set of 60 cancer cells.   

  Typically, protein-protein (gene-gene) interaction networks are modeled by association graphs, with nodes corresponding to proteins (genes), that has a single attribute, that is, protein profile (gene expression), and edges indicating some level of association between a pair of proteins (genes). Associations between pairs of proteins can indicate either direct binding and indirect participation in the same metabolic pathways or cellular process, and usually are known or inferred from corresponding protein profiles summarized into some association measure. Similarly, gene-gene associations may refer to direct co-regulation or indirect interaction in the same functional processes, and may also be known or inferred. Various measures of association have been used in the literature for the inference of biological association networks, including Pearson's product moment correlation (e.g., ~\cite{Steuer_2003}), partial correlation (e.g., ~\cite{Shipley_2002, Fuente_2004}), 
and mutual information (e.g., ~\cite{Butte1_2000,Butte2_2000,CLR_2007}).  See~\cite{Gardner_Faith_2005,Lee_2009}, for example, for reviews of association measures and their corresponding computational methods, as used in the context of inference of gene expression
networks.

As described in detail in Section~\ref{sec:methods}, we use correlation-based measures of association in this paper, that is, Pearson product moment correlations for networks based on individual attributes and canonical correlation for multi-attribute networks.  Although certainly the work of other authors has involved multiple types of data when inferring genomic networks 
(e.g., \cite{Shankavaram2007,Naylor_2010,Yamanishi_2003,Waaijenborg_2008}), to the best of our knowledge our work is the first to do so in a manner focused specifically on the notion of a multi-attribute network and its relation to the corresponding individual-attribute networks.

 By way of  illustration, consider the example of a simple protein network consisting of three nodes: Annexin A1, Annexin A2, and Keratin 8.  Annexin A1 and Annexin A2 are two calcium-binding proteins that are encoded by genes ANXA1 and ANXA2, respectively. Keratin 8 is a keratin protein encoded by the gene KRT8. Keratin 8 can be used to differentiate lobular carcinoma of the breast from ductal carcinoma of the breast. Annexin A1 has been of interest for use as a potential antiflamatory and  anticancer drug. The gene for Annexin A1 (ANXA1) is upregulated in hairy cell leukemia and can be used for diagnosing the disease. Annexin A2 is a less explored protein that is usually involved in the motility of the epithelial (skin) cells. 

\begin{figure}[h!]
\centering
\includegraphics[width = 1.6in, height = 1.3in] {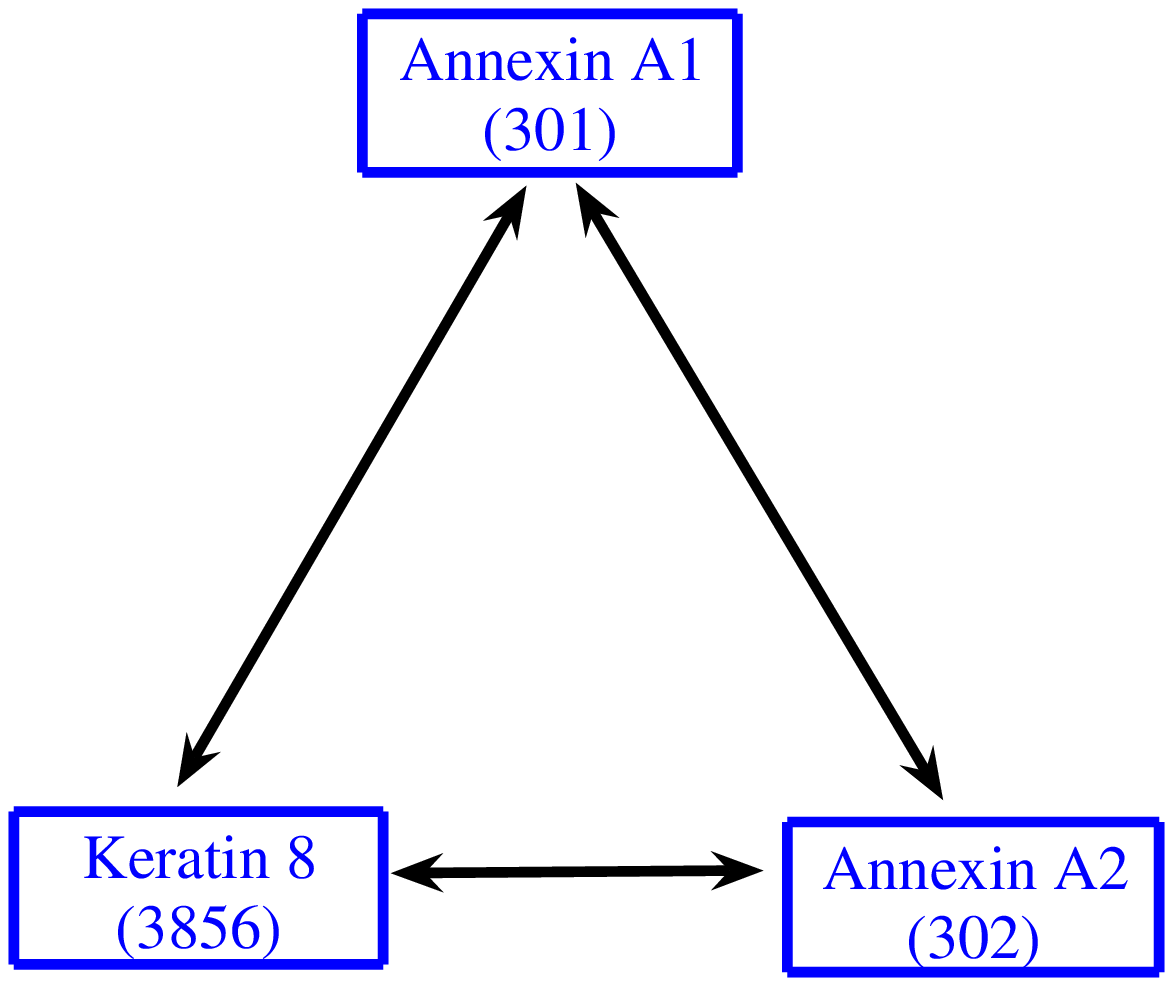}
\includegraphics[width = 1.6in, height = 1.3in] {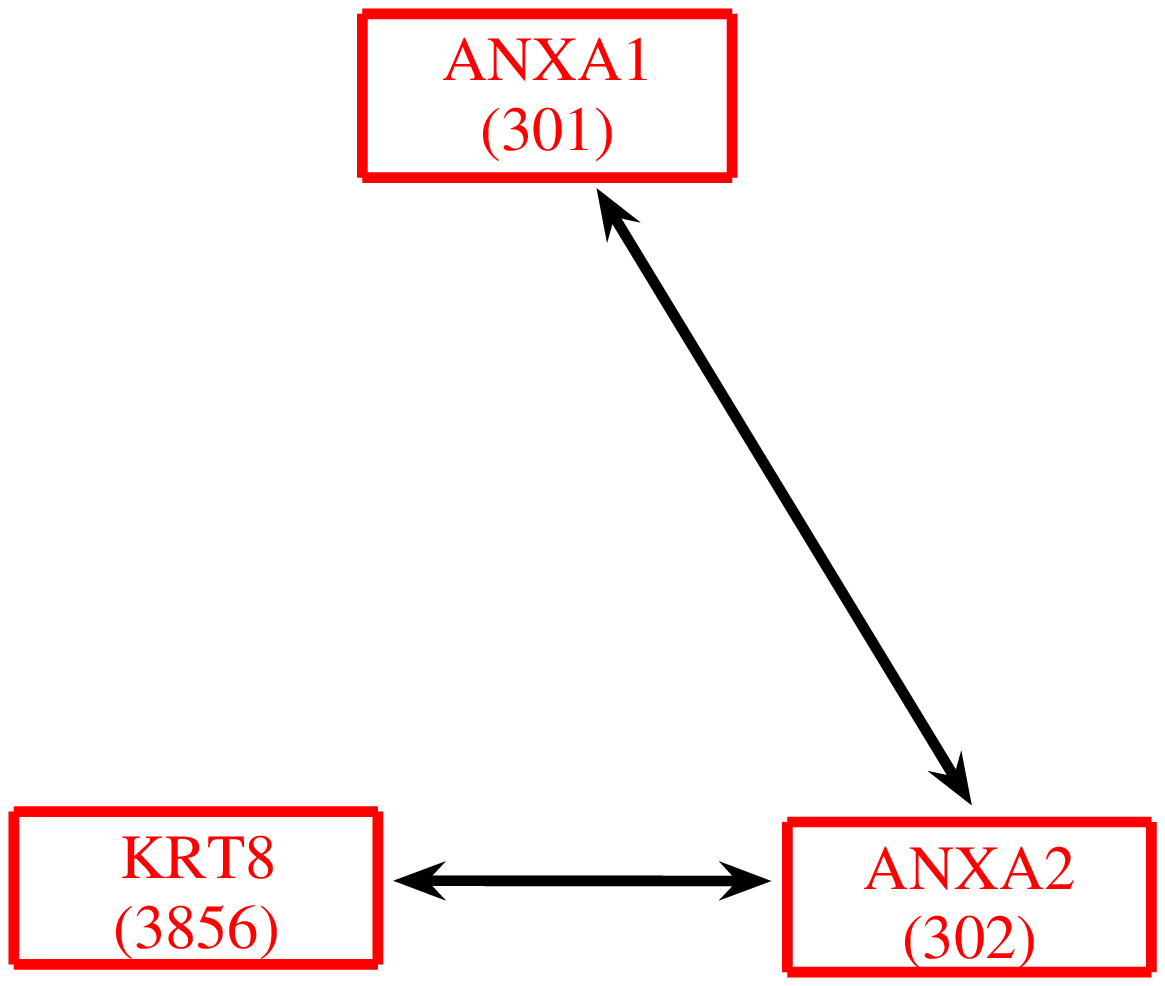}
\includegraphics[width = 1.6in, height = 1.2in] {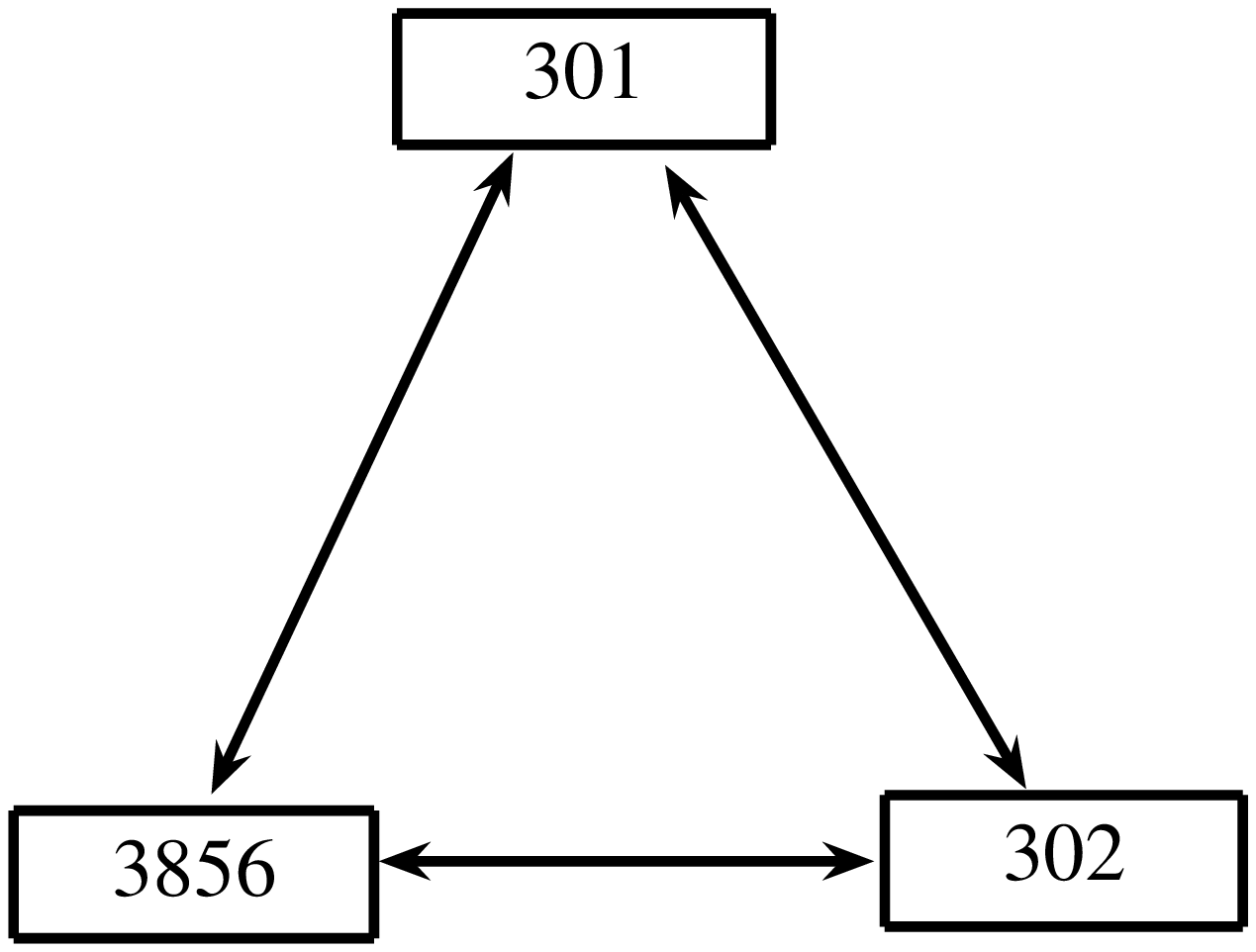}
\caption{Inferred association networks based on protein profiles (left panel), gene expressions (middle panel), and combined profiles (right panel). Numbers represent unique Entrez IDs.}
\label{fig:toynetworks}
\end{figure}

 Given protein profiles recorded on the same set of cells for all three nodes (Annexin A1, Annexin A2, and Keratin 8), we inferred the presence of links between all three pairs of nodes (left panel, Figure \ref{fig:toynetworks}); given corresponding gene expressions,  we inferred links only between ANXA2 and ANXA1 and between ANXA2 and KRT8 (middle panel, Figure \ref{fig:toynetworks}). This observation confirms the expectation that different molecular profiles can produce different networks, and,  hence, an association between protein profiles does not necessarily imply an association between corresponding gene expressions, and vice versa. A priori, it is not immediately clear how to compare these networks, and, more importantly, how to combine information based on both proteins profiles and gene expressions. 

 Motivated by these questions, we utilize the canonical correlation framework from classical  multivariate statistics to aggregate gene expression and protein profiles and construct a network based on combined profiles (right panel, Figure \ref{fig:toynetworks}). We see that the resulting network includes links between all three gene/protein pairs, like that network based only on protein profiles.  As we describe later, in the application of Section~\ref{sec:application}, we are also able to equip this network 
with numerical values summarizing the contribution of each type of data (i.e., protein profile versus gene expression) to each link, thus allowing us to offer an interpretation of the relative role of each link/node in this network in terms of gene and protein activity.  This interpretation may be used in turn to classify nodes (i.e., into proteomic, genomic, or `mixed' roles)
and we find, through enrichment analysis with a  biological databases on molecular pathways (i.e., KEGG\footnote{KEGG (Kyoto Encyclopedia of Genes and Genomes) is a bioinformatics resource for linking genomes to life and the environment, http://www.genome.jp/kegg/.}), that our classifications appear to be quite sensible when interpreted within the broader biological context.
%from the context of higher-order biological function.  
See Section~\ref{sec:application} for details.

\section{Multi-Attribute Association Networks}
\label{sec:methods}

By an {\it association network} we will mean a graph $G=(V,E)$, for nodes $v_i\in V$, $i=1,\ldots,N_v=|V|$, and edges $e(i,j)\in E$, in which edges indicate a sufficient level of association between the attributes of these nodes, according to some criterion function. Node attributes can be, for example, personal characteristics and preferences in social networks or levels of activity on different biological dimensions of a cell in biological networks.  Our interest here is in contexts where nodes are possessed of 
multiple attributes, all of which may enter into determining association between nodes.  That is, we are interested in 
{\em multi-attribute association networks}.  The main issue we consider in this section is the definition of a suitable summary of association between pairs of nodes and the relationship among such summaries when based on only
subsets of the full set of attributes. The question of inference of links in our network, given a choice of association measure, is addressed later in Section~\ref{sec:link_declaration}.

\subsection{Measures of Association}
\label{sec:similarity_measures}

 Suppose that for each node $i$ one can potentially observe $K$ attributes and  define a corresponding multivariate random variable  $X_i=(X^{(1)}_i, ..,X^{(K)}_i)^T$. In what follows, we assume that all attributes are continuous random variables.
 Let $SIM_C(i,j)$ be a specified measure of similarity between nodes $i$ and $j$ based on the subset of the node attributes $C$, where $C\subset \{1,..,K\}$. For a sufficiently `large' value of similarity $SIM_C(i,j)$ between nodes $i$ and $j$, an edge $e(i,j)$ will be assigned. In other words, we are interested in similarity measures $SIM_C(i,j)$ that constitute a 'nontrivial' level of association between attributes of two nodes $i$ and $j$ of network $G$. Usually, the similarity function $SIM_C(i,j)$ is not observable, but, nevertheless, can be potentially estimated from the information contained by measurements on  node attributes.       

Intuitively, it is expected that any chosen similarity measure $SIM_C(i,j)$ would differ for a different choice of subset of node attributes $C$. Therefore, it is important to understand how similarity measure $SIM_C(i,j)$ varies for different subsets of attributes within a given class of similarity measures. As a rule, the choice of an appropriate similarity measure, to a large extent, depends on a specific application. Here we restrict our attention to correlation-based similarity measures.

When only a single attribute is available ($K=1$), the Pearson product moment correlation 
\begin{equation}
 \rho(i,j)={\mathrm{cov}(X_i,X_j) \over \sqrt{ \mathrm{var}(X_i) \mathrm{var}(X_j)}}  
\label{eqn:pearsons_corr}
\end{equation}
is commonly used as a similarity measure.  When more than one node attribute is under consideration ($K>1$), Pearson's correlation between nodes $i$ and $j$ can be computed for each common attribute separately $\rho_l(i,j)=\mathrm{corr}(X_i^{(l)},X_j^{(l)}), ~ l\in C$, and then, if desired, computed values can be summarized into some aggregated measure of total between node similarity $SIM_C(i,j)$. For example:
\begin{itemize}
\item  Maximum correlation 
\begin{equation}
  SIM_C(i,j)\equiv \max_{l\in C} {\rho_l(i,j)} \enskip ,
\label{eq:max.corr}
\end{equation}
\item  Minimum correlation
\begin{equation}
  SIM_C(i,j)\equiv \min_{l\in C} {\rho_l(i,j)} \enskip .
\label{eq:min.corr}
\end{equation}
\end{itemize}

While these choices of multi-attribute similarity are intuitive and straightforward,  their main disadvantage is that they do not take into account the correlations between attributes observed on the same node and the cross-correlations between attributes observed on different nodes. 
From this perspective, canonical correlation is a more natural choice of total similarity for two main reasons. First, because it takes into consideration both the correlations between attributes on the same node and the cross-correlations between different attributes on different nodes, and second, because canonical correlation relates node sets of attributes in an optimal way. Additionally, canonical correlation analysis provides a way to evaluate the effective number and the importance of node attributes.  

Originally, introduced by \cite{Hotelling1936} and now a classical tool in multivariate statistics, we propose to use the canonical correlation $\rho_c(i,j)$ here as a measure of total similarity between multiple node attributes $X_i$  and $X_j$ of two nodes $i$ and $j$ in a network. We recall that computation of canonical correlation $\rho_c(i,j)$ is equivalent to maximization (in absolute value) of the correlation between two linear combinations $w_i^{T}X_i$ and $w_j^{T}X_j$ with respect to the vectors of weights $w_i\in \mathbb{R}^{|C|}$ and $w_j\in \mathbb{R}^{|C|}$, also called canonical weights:
\begin{equation}
 \rho_c(i,j) = \max_{w_i,w_j \in \mathbb{R}^{|C|}}\mathrm{corr}(w_i^{T}X_i, w_j^{T}X_j).
\label{eqn:cancorr_def}
\end{equation} 
 Note that since canonical weights $w_i$ and $w_j$ depend on a pair of indexes $(i,j)$, they are defined for each pair $(i,j)$ separately. However, we have suppressed this detail in our notation for the purpose of readability.
 
 By definition, the canonical correlation $\rho_c$ is a bounded quantity that takes values between zero and one.  By construction, $\rho_c$ is always greater or equal to the maximum in absolute value of any cross-attribute correlation between any pair of nodes in a network:   
$$ \rho_c(i,j) = \max_{w_i,w_j \in \mathbb{R}^{|C|}}\mbox{corr}(w_i^{T}X_i, w_j^{T}X_j) \ge \max_{l\neq k \in C} {|\mathrm{corr(X_i^{(l)}, X_j^{(k)} )} |}.$$

We will find it useful to adopt the eigenvalue formulation of the canonical correlation, and we will express this formulation in
terms of correlation matrices.
 Let $\Sigma_{ii}=\mathrm{Corr}(X_i)$ and $\Sigma_{jj}=\mathrm{Corr}(X_j)$ be the \textit{marginal} correlation matrices of attributes of node $i$ and node $j$, respectively;  and let $\Sigma_{ij}=\mathrm{Corr}(X_i, X_j)$ be the cross-correlation matrix between attributes of node $i$, and attributes of node $j$. Then the correlation supermatrix $\Sigma(i,j)$ can be represented as
 \begin{equation}
\Sigma(i,j) =\left(
\begin{array}{cc}
\Sigma_{ii}& \Sigma^{T}_{ij} \\
\Sigma_{ij} & \Sigma_{jj} 
\end{array} \right) \enskip ,
\end{equation}
and the canonical correlation \eqref{eqn:cancorr_def} can be expressed as
\begin{equation}
 \rho_c(i,j) = \max_{w_i,w_j \in {R}^{|C|}}\frac{w_i^{T}\Sigma_{ij} w_j}{\sqrt{w_i^{T}\Sigma_{ii} w_i}\sqrt{w_j^{T}\Sigma_{jj} w_j}},
\label{eqn:cancorr_opt} 
\end{equation}  
 where the vectors of weights $w_i$ and $w_j$ can be found directly by solving the optimization problem above, or by solving the system of eigenvalue equations
\begin{equation}
\left\{
\begin{array}{c}
\Sigma^{-1}_{jj} \Sigma_{ij}^{T} \Sigma^{-1}_{ii} \Sigma_{ij} w_j = \lambda^2 w_j \enskip  ,\\
\Sigma^{-1}_{ii} \Sigma_{ij} \Sigma^{-1}_{jj} \Sigma_{ij}^{T} w_i = \lambda^2 w_i \enskip .
\end{array}
\right.
\label{eqn:lagrange_system} 
\end{equation}
The canonical weights $w_i$ and $w_j$ are the eigenvectors that correspond to the maximum eigenvalue $\lambda^2$, the square root of which equals $\rho_c(i,j)$. 

Using canonical correlation, a natural criterion for assigning an edge between two nodes $i$ and $j$ is that $\rho_c(i,j)$ be greater than zero.  When an edge exists, the canonical weights $w_i, w_j$ and the canonical scores $w_i^T X_i, w_j^TX_j$ can be used to assess the relative contribution of each of the $K$ attributes to that edge.  This interpretation is an analogy to how we would evaluate the importance of explanatory variables in a multiple regression analysis.  Key ideas follow from the interpretation of these quantities. Specifically, the squared canonical correlation $\rho_c^2(i, j)$ is interpreted as the percentage of variation shared by the sets of attributes of nodes $i$ and $j$ along the directions defined by the canonical weights $w_i , w_j$.  Furthermore, the standardized canonical weights can be used to assess the relative importance of individual node attributes to a given canonical correlation. In particular, the squared, standardized canonical weight $(w_i^{(l)})^2, ~ l \in C,$ provides the relative contribution of attribute $l$ of node $i$ to $\rho_c(i, j)$. Finally, canonical scores $w_i^{T}X_i$ and $w_j^{T}X_j$ represent aggregated measures of attributes for nodes $i$ and $j$, respectively.

Often in network analysis  it is not unreasonable to assume a certain level of homogeneity across nodes in a network.  In the context of our model for multiple attributes, a natural set of homogeneity assumptions consists of assuming  (a) equality of the marginal correlation matrices, that is, $\Sigma_{ii}=\Sigma_{jj}$, and (b) symmetry of the cross-correlation matrix, that is, $\Sigma_{ij} = \Sigma_{ij}^T$.  The first assumption dictates that the correlations among attributes within a node are the same for both $i$ and $j$.  The second assumption dictates that the correlation among any pair of attributes between nodes $i$ and $j$, one from $i$ and one from $j$, respectively,  is unchanged if instead we look at those same two attributes but from $j$ and $i$.  In this case, we have the following result.
\begin{proposition}
Under the homogeneity assumptions that $\Sigma_{ii}=\Sigma_{jj}$ and $\Sigma_{ij}=\Sigma_{ij}^T$, 
the optimization \eqref{eqn:cancorr_opt} defining the canonical correlation
$\rho_c(i,j)$ between nodes $i$ and $j$ simplifies to
\begin{equation}
 \rho_c(i,j) =\max_{w\in \mathbb{R}^K}\frac{w^{T}\Sigma_{ij} w}{w^{T}\Sigma_{ii} w},
\label{eqn:cancorr} 
\end{equation}    
  and the corresponding eigenvalue problem is reduced to
\begin{eqnarray}
\Sigma^{-1}_{ii}~\Sigma_{ij} w = \lambda w.
\label{eqn:eigenproblem} 
\end{eqnarray}   
\label{prop:1}
\end{proposition} 
A proof of this result is given in the appendix.  This result has the important implication that, under homogeneity, only one set of canonical weights is required.  Therefore, when an edge exists between nodes $i$ and $j$, that is, when $\rho_c(i,j) > 0$, this single vector $w$ is a summary of the relative contribution of each attribute to the edge.  We will make use of this homogeneity assumption and the corresponding result  both in the illustration that follows next, in Section~\ref{sec:specialcase}, and in the simulations of Section~\ref{sec:simulation}.
In practice, these homogeneity conditions can be checked, for each pair $(i,j)$, using a simple likelihood ratio testing procedure, as we do in the application described in Section~\ref{sec:application}.

\subsection{Illustration: The Case of $K=2$}
\label{sec:specialcase}

 For the purpose of  illustration, we consider the special case of a single pair of nodes and $K=2$ attributes observed on each node. Let $X_i=(X^{(1)}_i, X^{(2)}_i)^T$ and $X_j=(X^{(1)}_j, X^{(2)}_j)^T$ be the attribute vectors for two nodes $i$ and $j$, with common marginal correlation matrix  $\mathrm{Corr}(X)\equiv \Sigma_m$ and symmetric cross-correlation matrix $\mathrm{Corr}(X_i,X_j)=\mathrm{Corr}(X_j,X_i)\equiv\Sigma_c$.  We parametrize $\Sigma_m$  and $\Sigma_c$ as 
\begin{equation}
\begin{array}{ccc}
\Sigma_m=\left(
\begin{array}{cc}
1 & r \\
r & 1 
\end{array} \right)
\,\hbox{and }\,
\Sigma_c=\left(
\begin{array}{cc}
\rho_1 & b  \\
b & \rho_2
\end{array} \right)
\, \hbox{,\, yielding}\quad 
\Sigma=\left(
\begin{array}{cc}
\Sigma_m & \Sigma_c  \\
\Sigma_c & \Sigma_m
\end{array} \right)
\end{array}
\enskip .
\nonumber
\end{equation}
\noindent Here the parameter $r=\mathrm{Corr}(X^{(1)},X^{(2)})$ represents the marginal correlation between the two attributes on a given node; $b=\mathrm{Corr}(X^{(1)}_i,X^{(2)}_j)=\mathrm{Corr}(X^{(2)}_i,X^{(1)}_j)$ is the cross-attribute correlation between nodes; and $\rho_1 = \mathrm{Corr}(X^{(1)}_i,X^{(1)}_j) $ and $\rho_2 = \mathrm{Corr}(X^{(2)}_i,X^{(2)}_j)$ are the within-attribute correlations between nodes for the first and the second attributes, respectively.    

To explore the space of parameter values where the canonical correlation $\rho_c$ is well-defined, and the effect of those parameter values on the value of $\rho_c$, we investigate the conditions under which the correlation matrix $\Sigma$ is positive-definite.  The  eigenvalues corresponding to $\Sigma$  are of the form
\begin{eqnarray*}
 %\mbox{eig}_{1,2}(\Sigma_m) &=& 1\pm r,\\
 %\mbox{eig}_{1,2}(\Sigma_c) &=& \frac{1}{2}\left(a_1+a_2\pm\sqrt{(a_1-a_2)^2+4b^2}\right).
 \mbox{eig}_{1,2}(\Sigma) &=& 1- \frac{(\rho_1 + \rho_2) \pm \sqrt{(\rho_1 - \rho_2)^2+4(b-r)^2}}{2},\\ 
 \mbox{eig}_{3,4}(\Sigma) &=& 1+ \frac{(\rho_1 + \rho_2) \pm \sqrt{(\rho_1 - \rho_2)^2+4(b+r)^2}}{2} \enskip .
 \end{eqnarray*} 
These eigenvalues are positive, and, consequently, $\Sigma$ is positive-definite, if the following conditions are satisfied:
\begin{equation}
\left\{ \begin{array}{l}
% |r|< 1, |a_1|<1, |a_2| < 1 , \\
 |b-r|< A_1=\sqrt{(1 - \rho_1)(1 - \rho_2)} \enskip ,\\
 |b+r|< A_2=\sqrt{(1 + \rho_1)(1 + \rho_2)}  \enskip .
\end{array}
\right.
\label{eqn:conditions}
\end{equation}
 
 The domain of the canonical correlation $\rho_c$ in terms of values $r$, $b$, for fixed values of  $\rho_1$ and $\rho_2$, where $\rho_2>\rho_1$, represents an oblique parallelepiped centered at the origin and with its size defined by values of $2A_1$ and $2A_2$, which in turn depend on $\rho_1$ and $\rho_2$. The corresponding value of the canonical correlation can be computed explicitly by solving $\Sigma_m^{-1}\Sigma_c x = \lambda x$ with respect to $\lambda$, yielding
\begin{eqnarray}
\rho_c=\max\left\{\left\vert\mbox{eig}_{1,2}(\Sigma_m^{-1}\Sigma_c)\right\vert\right\} &=& \max\left\{\left\vert
\frac {\rho_1+\rho_2-2br \mp\sqrt {D}}{2(1-r^{2})}\right\vert \right\},\\
\mbox{where   } ~ D &=&(\rho_1 - \rho_2)^2 +4(b-\rho_1 r)(b-\rho_2 r)\nonumber \enskip .
\end{eqnarray}

\begin{figure}[!htb]
 \centering
 \caption{Domain of canonical correlation (left panel) and actual values of canonical correlation (right panel) computed for fixed values of $\rho_1=0.3$ and $\rho_2=0.1$ as functions of $r$ and $b$.}
\includegraphics[width=2.2in, height=2.0in]{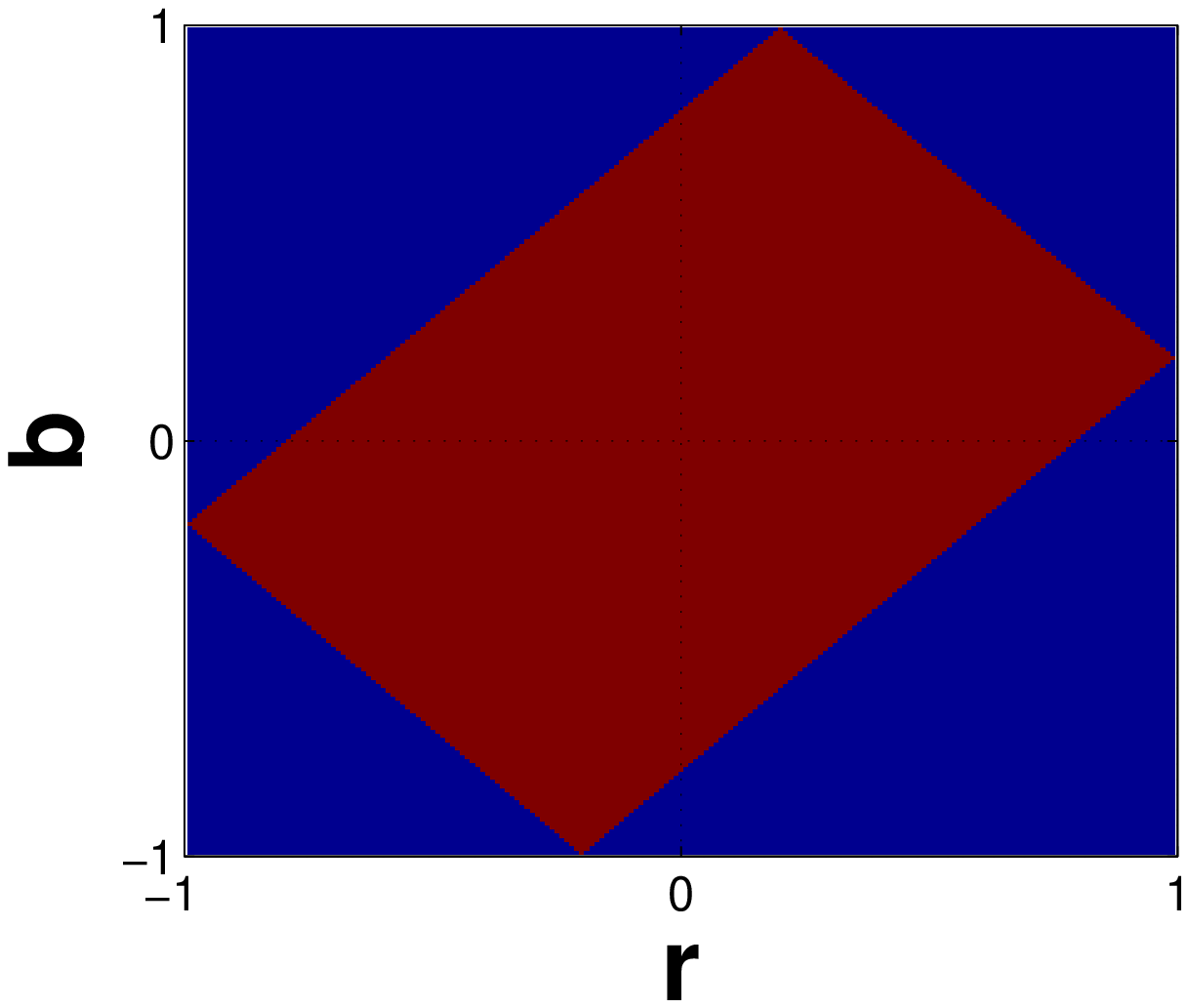} \hskip 0.01in 
\includegraphics[width=2.2in, height=2.0in]{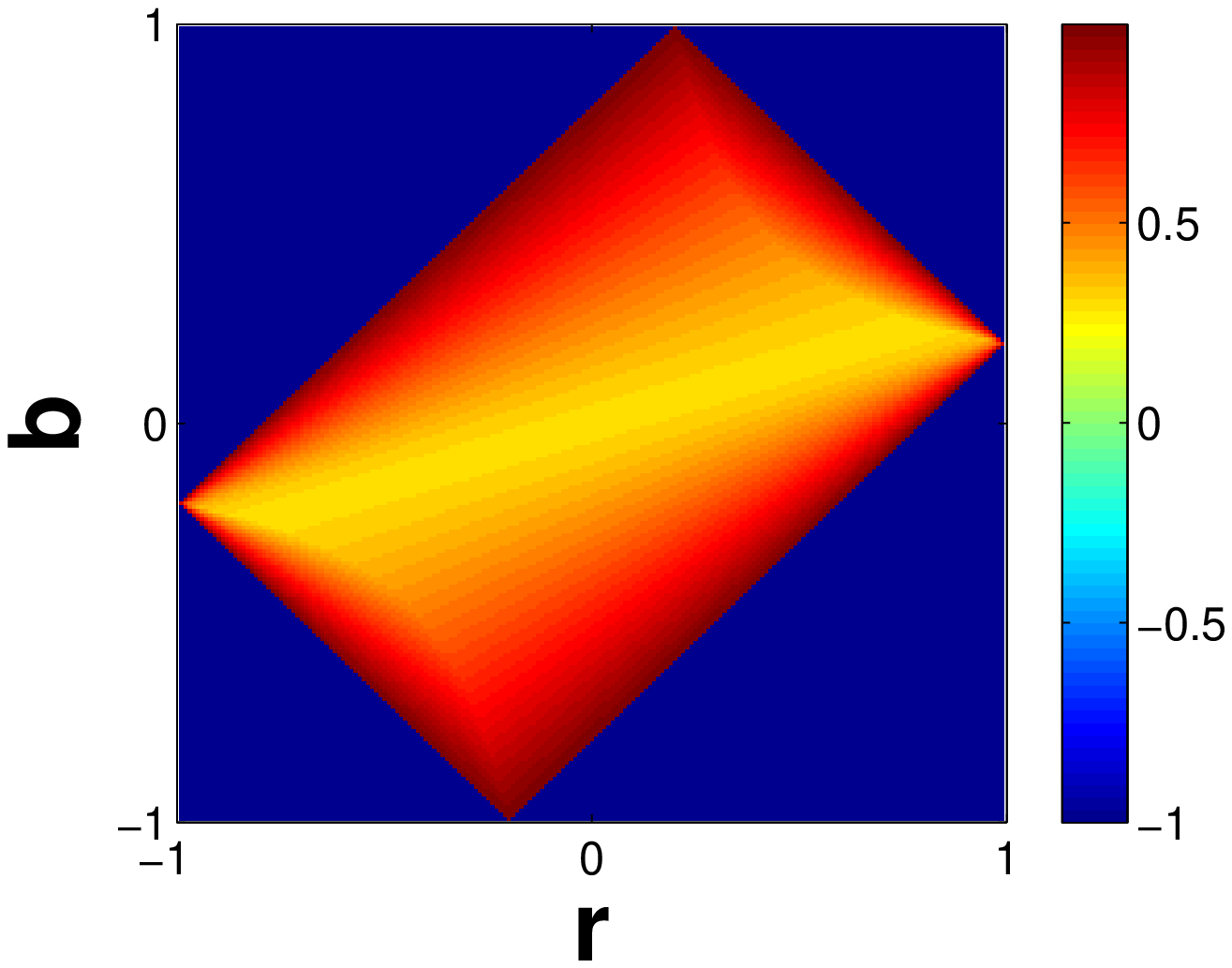}
 \label{fig:rho_area}
 \end{figure}

\begin{figure}[!htb]
 \centering
 \caption{Squared standardized canonical weights $w_1^2$ (left panel) and $w_2^2$ (right panel) computed for fixed values of $\rho_1=0.3$ and $\rho_2=0.1$ as functions of $r$ and $b$.}
\includegraphics[width=2.2in, height=2.0in]{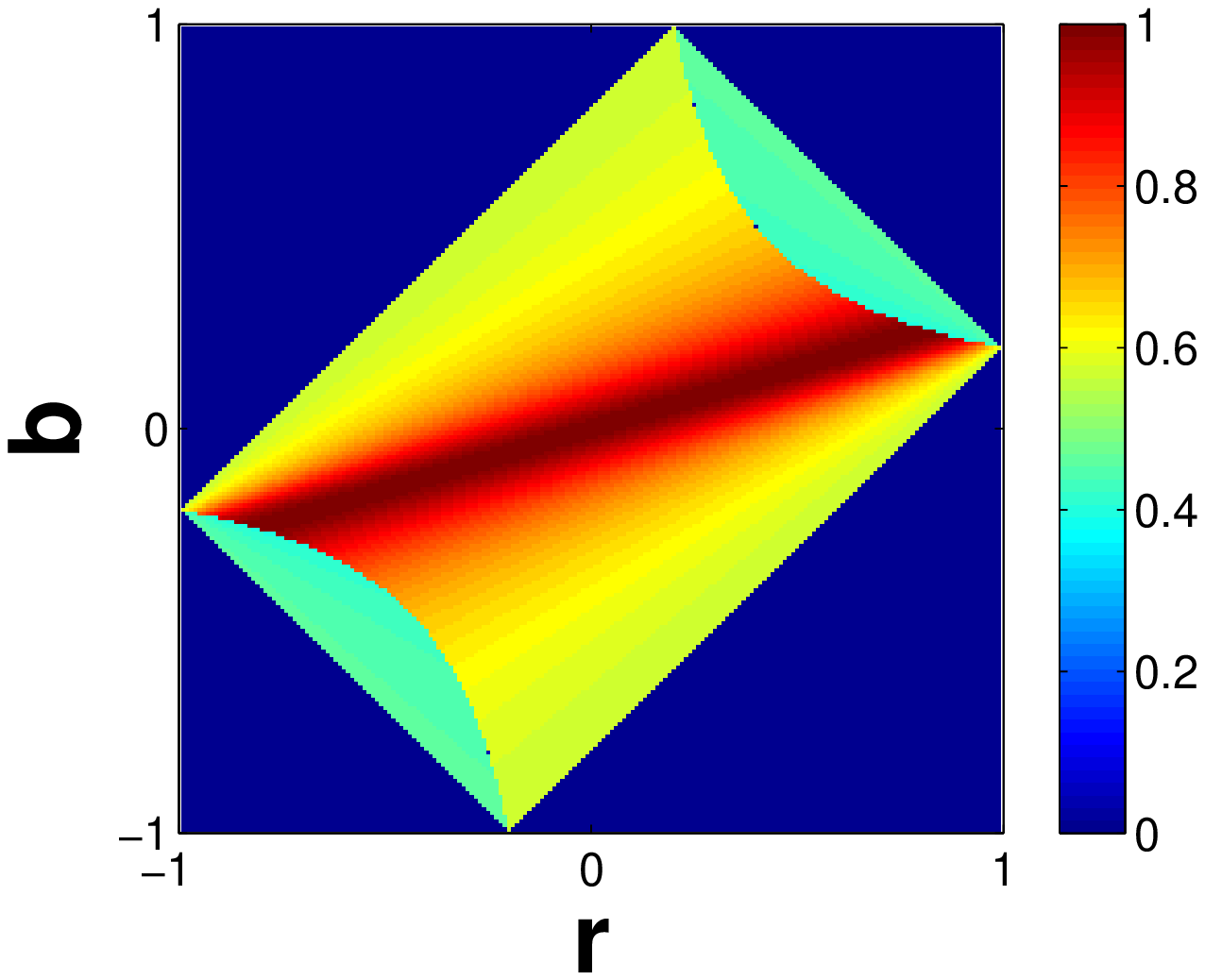} \hskip 0.01in 
\includegraphics[width=2.2in, height=2.0in]{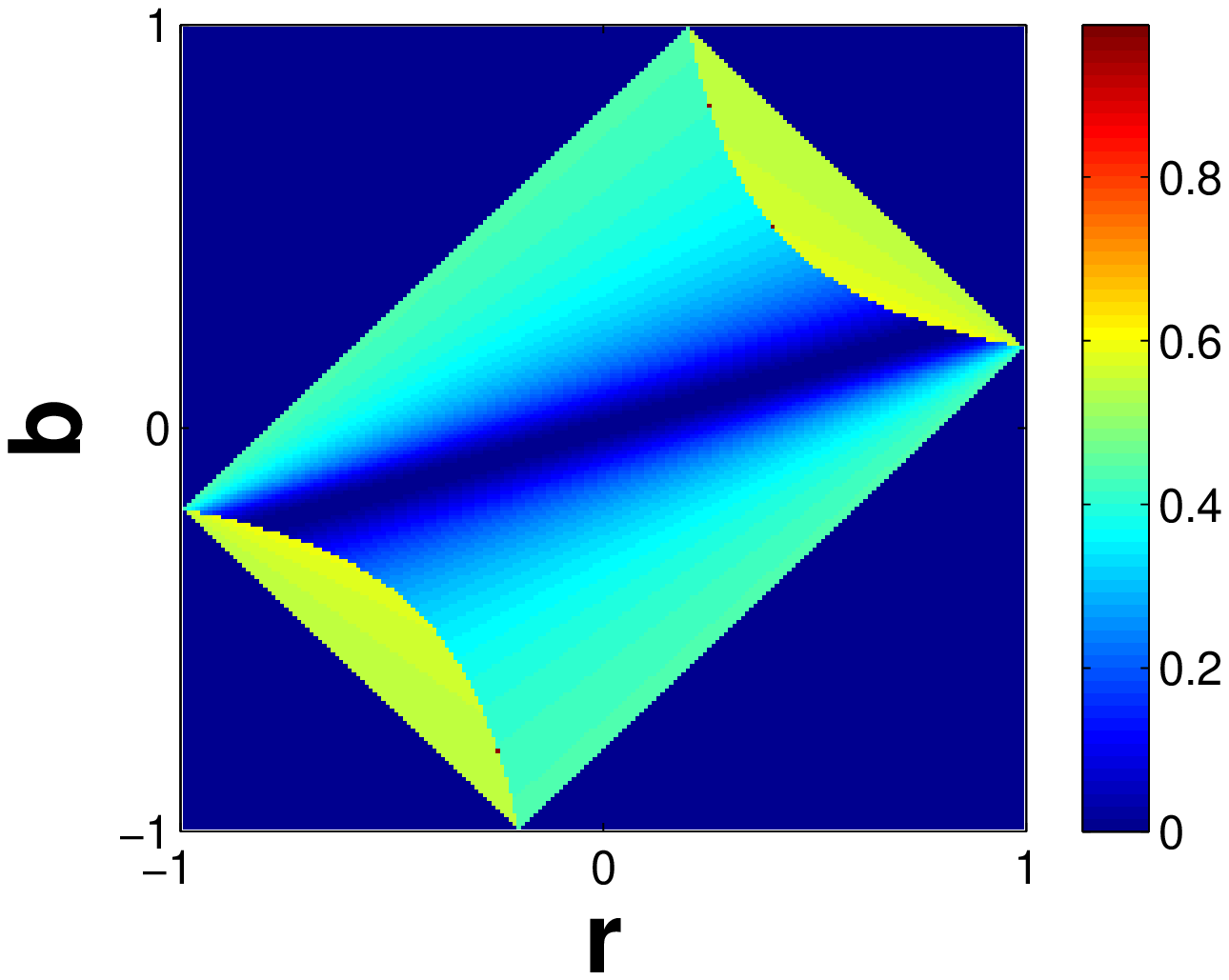} 
 \label{fig:rho_weights}
 \end{figure}

 Figure \ref{fig:rho_area} shows the domain of canonical correlation (left panel) and actual values of canonical correlation (right panel) computed for fixed values of $\rho_1=0.3$ and $\rho_2=0.1$ as functions of $r$ and $b$. If the cross-correlation $b$ is induced by correlation $r$ between attributes of the same node, then the canonical correlation is not noticeably greater than the maximum in absolute value of $\rho_1$, $\rho_2$, and $b$. However, if substantial cross-correlation $b$ exists between different attributes, then the value of the canonical correlation is noticeably greater than $\rho_1$, $\rho_2$, or $b$.      
 Canonical weights are depicted in Figure \ref{fig:rho_weights}. Since all necessary conditions of Proposition~\ref{prop:1} are satisfied, only one set of weights $(w_1,~w_2)$ for each pair of nodes needs to be computed. Squared, standardized weights $w_1^2$ and $w_2^2$, in this scenario, provide relative contribution of the first and the second attributes to $\rho_c$. When $b$ is relatively small, meaning, there is no substantial cross-correlation between different attributes of different nodes, the value of canonical correlation is effected, to a large extent, by that  attribute on which the correlation between two nodes is the strongest. This results in a large value of $w_1^2$ (close to one), and consequently a small value of $w_2^2$ (close to zero). For small and moderate values of $r$, as the cross-correlation increases in absolute value, the value of canonical correlation also increases, and so too the influence of the second attribute. This tendency results in lower values of $w_1^2$ and higher values of $w_2^2$. Due to the constraints on $r$ and $b$ for obtaining a valid covariance matrix $\Sigma$, not all combinations of these parameters result in proper values of $\rho_c$, $w_1$, and $w_2$.
  
For $K>2$, in the simplest scenario, where all off-diagonal elements of the matrix $\Sigma_m$ are equal to $r$ and all diagonal elements equal to $1$, and all off-diagonal elements of the cross-covariance matrix $\Sigma_c$ are equal to $b$ and diagonal elements equal to $\rho$, the corresponding eigenvalues of $\Sigma$ can be computed explicitly:
\begin{eqnarray*}
%\mbox{eig}_{1,..,(k-1)}(\Sigma_m) &=& 1 - r \nonumber, ~~ \mbox{eig}_{k}(\Sigma_m) = 1 + (k-1)r \nonumber\\
\mbox{eig}_{1,2,..,(k-2)}(\Sigma )&=&
 (1 - r) \pm (\rho - b) \nonumber, \\
 \mbox{eig}_{(k-1),k}(\Sigma )&=&(1 + (k-1)r) \pm (\rho+(k-1)b) \nonumber  \enskip .
\end{eqnarray*}
These values are positive provided
$$  -1/(k-1)<r<1, ~~~|\rho-b|<|1-r|, ~~~ |\rho+(k-1)b|<|1+(k-1)r| ,$$
and the corresponding canonical correlation is
$$
\rho_c=\max \left\{ \left|\frac{\rho-b}{1 - r} \right|,\left|\frac{\rho+(k-1)b}{1 +(k-1)r}\right| \right\}.
$$
 In this situation, there are only two unique canonical roots, and so we can use any two or even one attribute to infer links in the network. In general, however, for networks with an arbitrary number $K$ of multiple attributes per node and less trivial correlation structure, the number of parameters increases significantly, so that an explicit expression of the canonical correlation becomes intractable. 
%
%%%%%%%%%%%%%%%%%%%%%%%%%%%%%%%%%%%%%%%%%%%%%%%%%%%%%%%%%%%%%%%%%%%%%%%%%%%
\section{Network Topology Inference} 
\label{sec:link_declaration}

We describe here a testing-based approach to inferring multi-attribute association networks and we present 
the results of a small simulation study comparing the power of edge detection using the several definitions of similarity
detailed above in the previous section.

\subsection{Methods}
\label{sec:methods_of_edge_detection}

Recall that a link between two nodes $i$ and $j$ in a multi-attribute association network $G=(V,E)$ is  present when there is sufficient similarity $SIM_C(i,j)$ between the corresponding sets of attributes $X_i$ and $X_j$, based on some choice of subset $C\subseteq \{1,\ldots,K\}$ of $|C|$ attributes.  Given appropriate data, we wish to infer the topology of our network $G$.  In general, for inference of single-attribute association networks methods are of two types: those based on principles of hypothesis testing and those based on regression principles.  See~\citet[Chap. 7.3]{Kolaczyk} for an overview.  Here we choose to employ a testing-based approach for inferring multi-attribute association networks.

Specifically, given a choice of similarity $SIM_C(i,j)$, and $n$ independent and identically distributed observations $\{(x_{ik},x_{jk})\}_{k=1}^n$ of the random variable pair $(X_i,X_j)$ of attributes for a pair of nodes $i$ and $j$, we approach the task of determining whether $e(i,j)\in E$ as one of testing the hypotheses
\begin{equation}
H_0: ~ SIM_C(i,j) = 0 ~~ \mbox{  versus  } H_1: ~ SIM_C(i,j)\neq 0  \enskip .
\end{equation}
We test each such pair of nodes $(i,j)$, for $i,j\in V$ and $i< j$, and control for the large number of tests (i.e., $N_v(N_v-1)/2$ in all)
using false discovery rate principles, through application of the method of ~\cite{Benjamini_1995}.

The network $G$ of primary interest to us in this paper is that defined through the use of canonical correlation as our similarity measure.  The corresponding hypothesis testing problem is
\begin{eqnarray}
H_0: ~ \rho_c(i,j) = 0 ~~ \mbox{  versus  } H_1: ~\rho_c(i,j)\neq 0 \enskip .
\end{eqnarray}
There are several test statistics from classical multivariate statistics that can be used in testing these hypotheses.   Here we employ the one arguably most commonly used,  Bartlett's $\chi^2$ statistics \citep{Bartlett_1941}.  Specifically, we compute for each pair $(i,j)$ the statistic
\begin{equation}
    \chi ^2(i,j) = - \left[ (n -1) - (|C| + 0.5) \right] \ln \prod _ {l = 1} ^{|C|} \left[ 1 - \hat{\rho}_{c(l)}^2(i,j)\right] \enskip ,
\label{eqn:Bartlett}
\end{equation}
which, by Wilk's theorem, under $H_0$ is asymptotically distributed as a $\chi^2$ random variable with $|C|^2$ degrees of freedom, when applied to a subset $C\subseteq \{1,\ldots,K\}$ of $|C|$ attributes.  Note that in order to compute this statistic it is necessary to estimate the marginal and cross-correlation matrices for each edge $i$ and $j$ and to solve the generalized eigenvalue problem
\eqref{eqn:cancorr_opt} (or, under homogeneity, the eigenvalue problem \eqref{eqn:eigenproblem}),  computing all eigenvalue roots $\hat{\rho}_{c(l)}^2 = \lambda_l, l = 1, ... , |C|$. This may be done using standard software. In addition, in order to estimate $(2|C|)$ dimensional super-correlation matrix,  for each attribute, one needs to have at least $(2|C|)(2|C|-1)/2$  independent observations.  In the absence of sufficiently large numbers of observations, if the underlying network is expected to be sufficiently sparse, an alternative would be to compare the test statistic to a null distribution derived from empirical null principles \citep{Efron_2010}.

 Note that by declaring an edge based on Bartlett's $\chi^2$ statistics \eqref{eqn:Bartlett}, we use canonical variables of all orders $\rho_{c(l)}^2 = \lambda_l, l = 1, ... , |C|$ by definition. However, once an edge is declared, we assign it canonical weights that correspond to the first order (the maximum)canonical correlation $\rho_{c}\equiv\rho_{c(1)}$.

By way of comparison, and in preparation for our simulation study below, we also  consider the corresponding testing procedures for  inference of $G$ based on (i) just a single attribute and Pearson's product moment correlation, and (ii) a max- or min-based aggregation across attributes, combining the individual Pearson correlations per the expressions in \eqref{eq:max.corr} and~\eqref{eq:min.corr}. 

 In the case where only a single attribute is used for each node (indeed, perhaps only a single attribute is observed), and Pearson's correlation is used as a measure of similarity between a pair of nodes, a link between nodes $i$ and $j$ is declared according to the following test of hypotheses:
\begin{equation}
H_0: ~ \rho(i,j) = 0 ~~ \mbox{  versus  } H_1: ~ \rho(i,j)\neq 0 \enskip .
\end{equation}
The natural test statistic is the empirical correlation $\hat\rho(i,j)$, which is commonly transformed and compared to either standard 
normal distribution or an appropriate Student's $t$-distribution.  See~\citet[Chap. 7.3.1]{Kolaczyk}.  Here we adopt the former formulation, based on Fisher's transformation, comparing the statistic
\begin{equation}
z(i,j)= \frac{\sqrt{n-3}}{2}\, \log \left\{ \frac{(1+\hat{\rho}(i,j))}{(1-\hat{\rho}(i,j))} \right\}, 
\label{eqn:Fishers}
\end{equation}
to a normal distribution with mean zero and variance one.

In the case of max- or min-based aggregation across attributes, a link between nodes $i$ and $j$ is declared according to the
following tests of hypotheses, respectively:
\begin{eqnarray}
H_0: ~ \rho_l(i,j) = 0,~ \forall  l\in C ~~ \mbox{  versus  } H_1: ~ \max_{l\in C}{\rho_l(i,j)}\neq 0\\
H_0: ~\rho_l(i,j) = 0,~ \forall  l\in C  ~~ \mbox{  versus  } H_1: ~ \min_{l\in C}{\rho_l(i,j)}\neq 0 \enskip . \nonumber
\end{eqnarray}
Here, we estimate the sample correlation $\hat{\rho}_l(i,j)$ for each attribute $l \in C$ and compute the corresponding testing statistic $z_l(i,j)$ using Fisher's transformation \eqref{eqn:Fishers}. Since $z(i,j)$ is an increasing function of $\hat{\rho}(i,j)$, the maximum (minimum) of $z_l(i,j)$ will correspond to the maximum (minimum) of $\hat{\rho}_l(i,j)$. To calculate $p$-values associated with such tests,  approximations based on the so-called rhombus formula may be used~\cite{Efron_1997, Li_2008}.

%%%%%%%%%%%%%%%%%%%%%%%%%%%%%%%%%%% K=2 %%%%%%%%%%%%%%%%%%%%%%%%%%%%%%%%%%%%%%
\subsection{A Simulation Study}
\label{sec:simulation}

 In order to gain some insight into the comparative behavior of these different test-based approaches to inferring association
networks, and the different ways in which they utilize information on multiple attributes, 
we conducted a small simulation study.  In what follows we evaluate numerically the power of each test to infer
an individual link.  Specifically, we infer the presence of a link defined through (1) Pearson's correlation measured on the first attribute, based
on $\rho_1>0$; (2) Pearson's correlation measured on the second attribute, based on $\rho_2 > 0$; (3) the maximum correlation, $\max(\rho_1,\rho_2) > 0$; (4) the minimum correlation, $\min(\rho_1,\rho_2) > 0$; and (5) the canonical correlation, $\rho_c$.
The corresponding hypotheses to be tested are
\begin{enumerate}
\item $H_0:~ \rho_1 = 0 \mbox{ vs. }  H_1:~ \rho_1 > 0$, 
\item $H_0:~ \rho_2 = 0 \mbox{ vs. }  H_1:~ \rho_2 > 0$, 
\item $H_0:~ \rho_1 =\rho_2 = 0 \mbox{ vs. }  H_1:~ \max(\rho_1,\rho_2) > 0, ~ (\rho_1 >0 \mbox{ or } \rho_2 >0 )$, 
\item $H_0:~ \rho_1 =\rho_2 = 0 \mbox{ vs. }  H_1:~ \min(\rho_1,\rho_2) > 0, ~ (\rho_1 >0 \mbox{ and } \rho_2 >0 )$,
\item $H_0:~ \rho_c = 0 \mbox{ vs. }  H_1:~ \rho_c > 0$. 
\end{enumerate}
    
Our simulations are performed under the following setup. We fix values $\rho_1$ and $\rho_2$ to be $0.3$ and $0.1$, respectively and generate $1000$ independent data samples of size $n=50$ from the multivariate normal distribution $(X,Y)\sim N_{4}(0,\Sigma)$, where $\Sigma$ is defined as in Section~\ref{sec:specialcase}, over a range of values for $r$ and $b$. Given simulated data, we estimate the values of $\rho_1$, $\rho_2$,  and $\rho_c$ and compute the appropriate test statistics, as described in Section~\ref{sec:methods_of_edge_detection}, and evaluate the power of the tests under the described five sets of hypotheses. For Scenario 3, we approximate p-values using a simplified version of the rhombus formula, the so-called W-formula, derived by \cite{Efron_1997} and fitted for $k=2$: 
\begin{eqnarray}
%\Pr(\max(\hat{\rho}_1(i,j),\hat{\rho}_2(i,j)) > c) = 
\Pr(\max(z_1(i,j),z_2(i,j)) >c) \approx \bar{\Phi}(c) + \phi(c)\frac{\phi(cL/2)-0.5}{c/2},  
\end{eqnarray}
where $L=\arccos(corr(z_1(i,j),z_2(i,j)))$ and $c$ is an observed value of the maximum of test statistics $z_1(i,j)$ and $z_2(i,j)$,
with $\bar{\Phi}$ and $\phi$ denoting the complementary cumulative distribution function and the density function
of the standard normal, respectively. Analogously, for Scenario 4, we have:
\begin{eqnarray}
%\Pr(\max(\hat{\rho}_1(i,j),\hat{\rho}_2(i,j)) > c) = 
\Pr(\min(z_1(i,j),z_2(i,j)) >\tilde{c}) \approx \bar{\Phi}(\tilde{c}) - \phi(\tilde{c})\frac{\phi(\tilde{c}L/2)-0.5}{\tilde{c}/2},  
\end{eqnarray}
where $\tilde{c}$ is an observed value of the maximum of test statistics $z_1(i,j)$ and $z_2(i,j)$.
Note that association exists (i.e., there is an edge present) under all five measures of similarity.

\begin{figure}[h!]
\centering
\includegraphics[width = 2.6in, height = 1.8in] {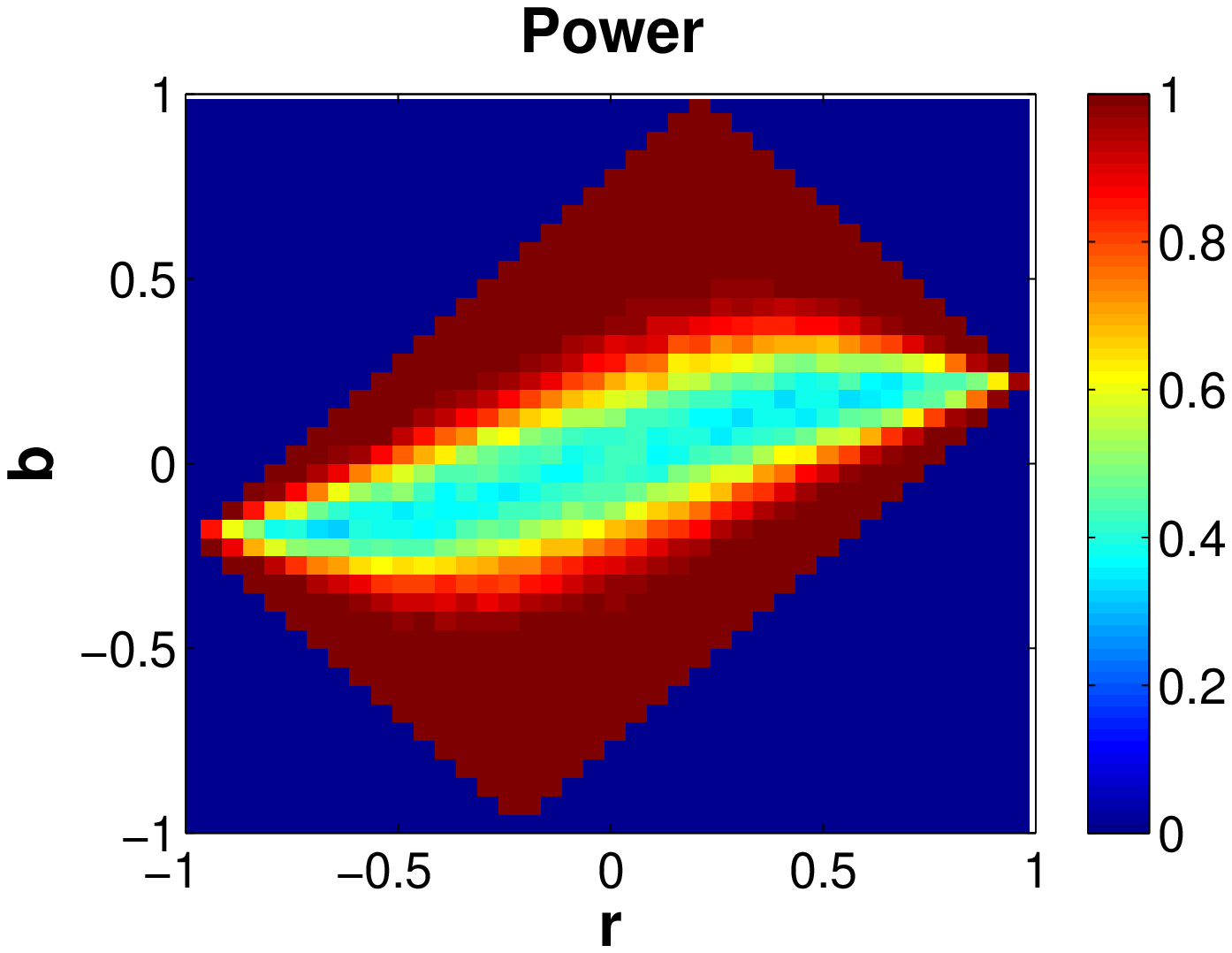} \vskip 0.01in 
\includegraphics[width = 2.4in, height = 1.8in] {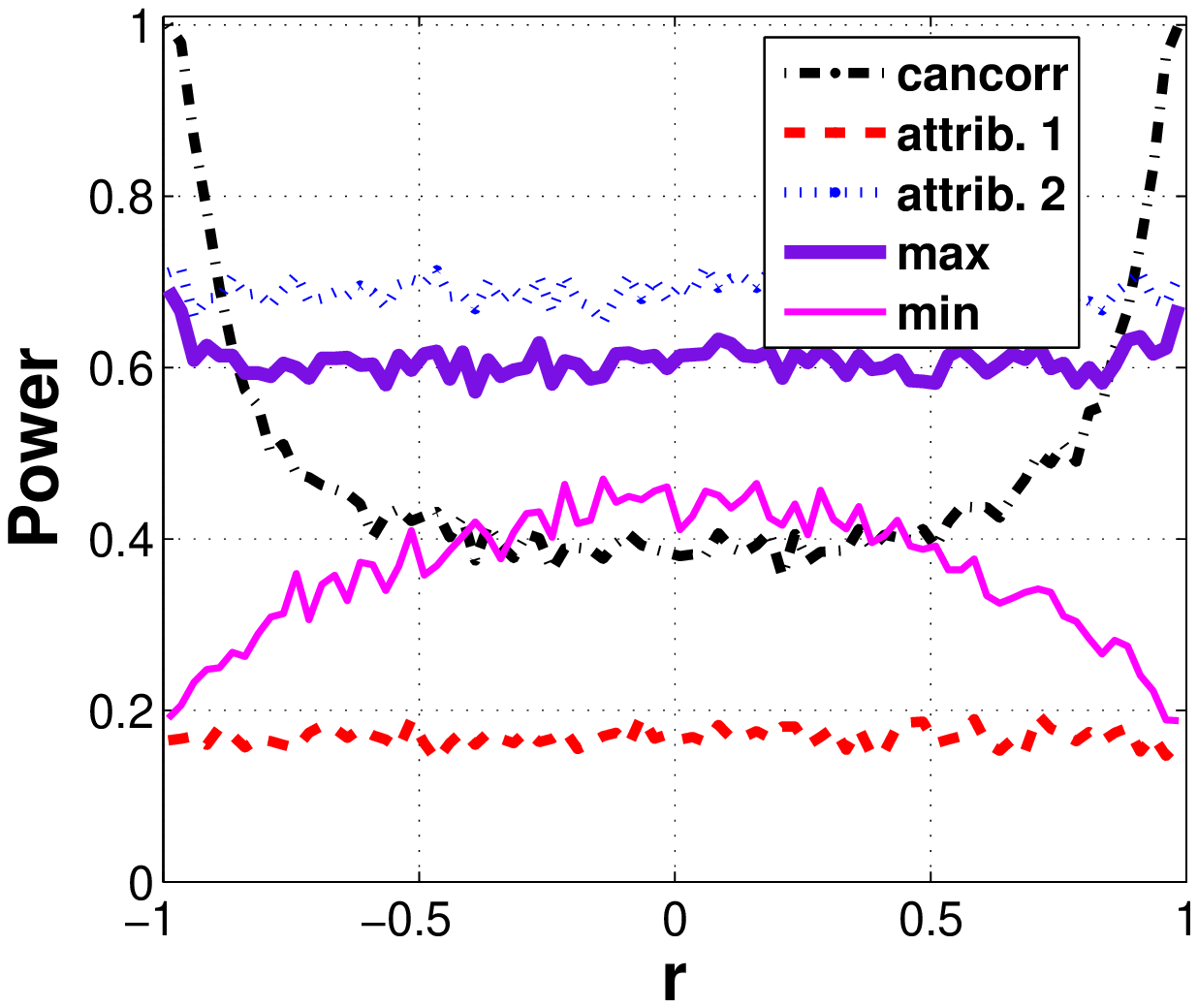} \hskip 0.01in
\includegraphics[width = 2.4in, height = 1.8in] {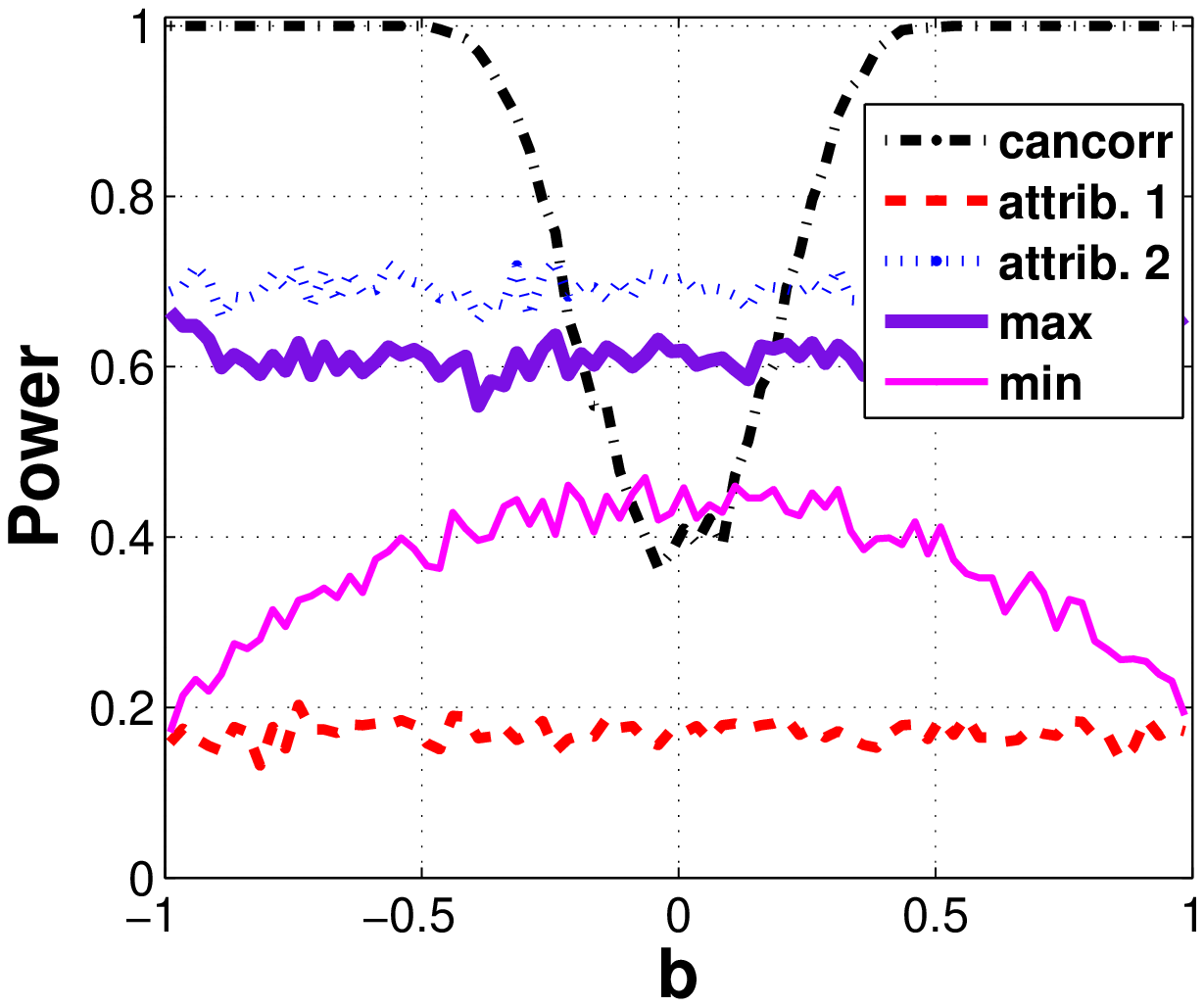}
\caption{Top panel shows power for canonical correlation only (scenario (5)); left and right panels present the power for all five described scenarios as a function of $r$ (where $b=0.2~r$) and as a function of $b$ (where $r = 0.2\,b$).}
\label{fig:power}
\end{figure}

 The results of the simulations are depicted in Figure \ref{fig:power}. The top panel of Figure \ref{fig:power} shows power as a function of $r$ and $b$ for canonical correlation only. Recall that $r$ is the correlation between attributes for a given vertex (i.e., within-vertex correlation), while $b$ is the correlation between attributes across two vertices (i.e.,  between-vertex correlation).  From the top panel in the figure it is clear that, while power increases as the within-vertex correlation $r$ increases, for a fixed value of $r$ even a small amount of between-vertex correlation $b$ is sufficient to greatly increase power.

Now consider the left and right panels of Figure \ref{fig:power}, in which we  present the power for all five described scenarios as a function of $r$ (where $b=0.2~r$) and as a function of $b$ (where $r = 0.2\,b$). The power curves for detecting the edge when using either the first or second attribute alone indicate what may be achieved with only partial information, {\it that is, on only one attribute or
the other.}  That the higher power curve corresponds to the first attribute is natural, given that $\rho_1=0.3 > 0.1 = \rho_2$.
More interestingly, we see that among the three scenarios under which information on both attributes is used, only that
based on canonical correlation of attributes is capable of exceeding the power using the first attribute alone.  More specifically,
the left  panel shows the situation where the within-vertex correlation $r$ varies from $-1$ to $1$, but at the same time cross-correlation between two nodes stays relatively small, in a range of $(-0.2,~ 0.2)$. In this case, the effect of the correlation based on the first attribute on the power of link detection is reduced, and hence the power of the test for canonical correlation  decreases. In contrast, when the cross-correlation between two nodes $b$ grows more rapidly than correlation $r$, the power of the test for canonical correlation  increases similarly rapidly and quickly achieves a maximum of $1.0$.

 Thus, by means of this small, illustrative simulation study, we were able to provide qualitative explanation of the  relationship between the power for detecting an edge under the five different scenarios and, in particular, gain some insight into the way in which differing extents to which information on multiple attributes is used can affect the power.

\section{Inference and Characterization of a Gene-Protein Network}
\label{sec:application}
 
 In this section, we turn our attention to the gene/protein regulatory network application introduced in Section \ref{sec:motivation}. We analyze a subset of the NCI-60 database that contains 92 protein profiles and gene expressions for approximately $9,000$ genes. Note that the problem of combining multiple types of biological profiles is nontrivial. We adopt the procedure described in~\cite{Shankavaram2007} to construct a so-called `concensus' data set comprised of 91 protein profiles and 91 gene profiles matched in corresponding pairs by their common gene/protein Entrez identifiers. In this manner we obtain a set of bivariate measurements on the expression for each 91 genes/proteins across 60 cancer cells.

\subsection{Network Inference and Characterization}
\label{sec:netcharacterization}

We inferred three types of networks: a network of associated proteins, based on similarity of  protein expression profiles alone; a network of associated genes, based on similarity of gene expression profiles alone; and a single gene/protein network, based on both types of expression profiles.  We used the methods of hypothesis testing described in Section~\ref{sec:link_declaration}, with an FDR control level of $\gamma=0.05$. Note that since we found (using formal hypothesis testing) that network homogeneity is not supported for all pairs of nodes in the gene-protein network, the simplified homogeneous covariance structure discussed in parts of Section~\ref{sec:methods} is not assumed here.

\begin{figure}[h!]
 \centering
 \caption{Inferred association network based on protein profiles (left panel), gene expressions (middle panel), and gene and protein profiles combined (right panel). Numbers in boxes represent unique Entrez IDs; numbers on edges represent estimated correlations and, for gene-protein network (right panel), and corresponding canonical weights. Dashed lines indicate absent edges.}
\includegraphics[width = 1.2in, height = 1.2in] {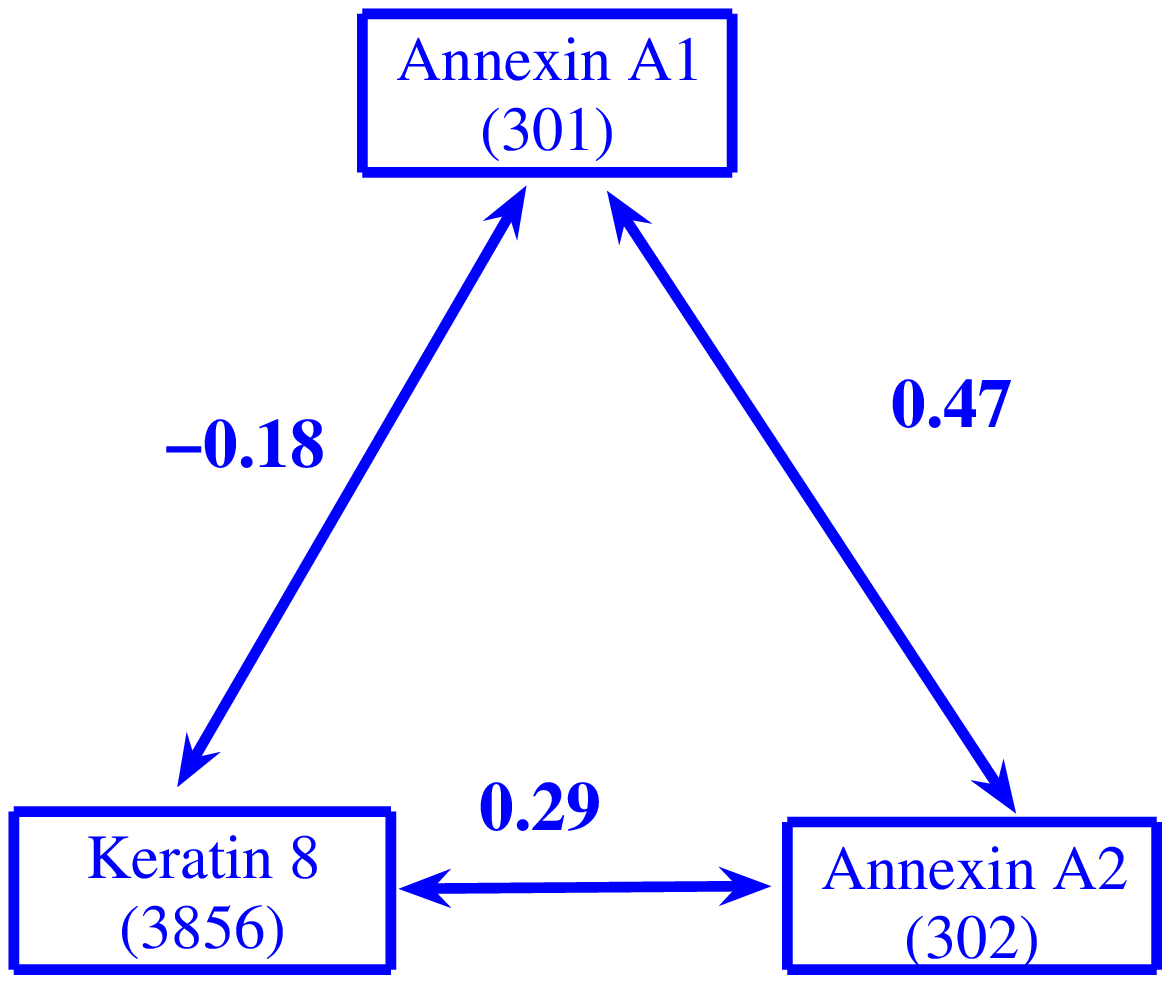}
\includegraphics[width = 1.2in, height = 1.2in] {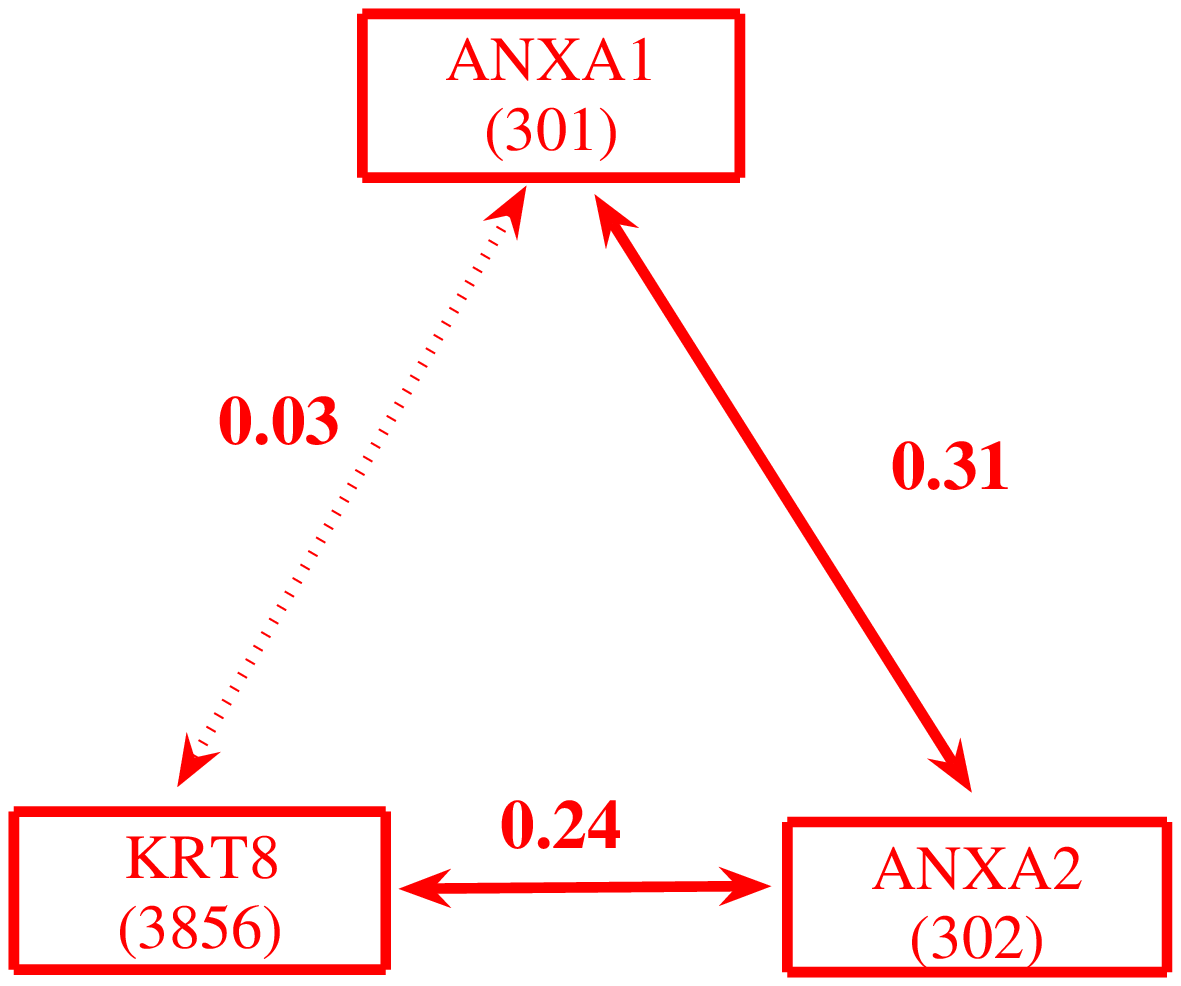}
\includegraphics[width = 1.9in, height = 1.0in] {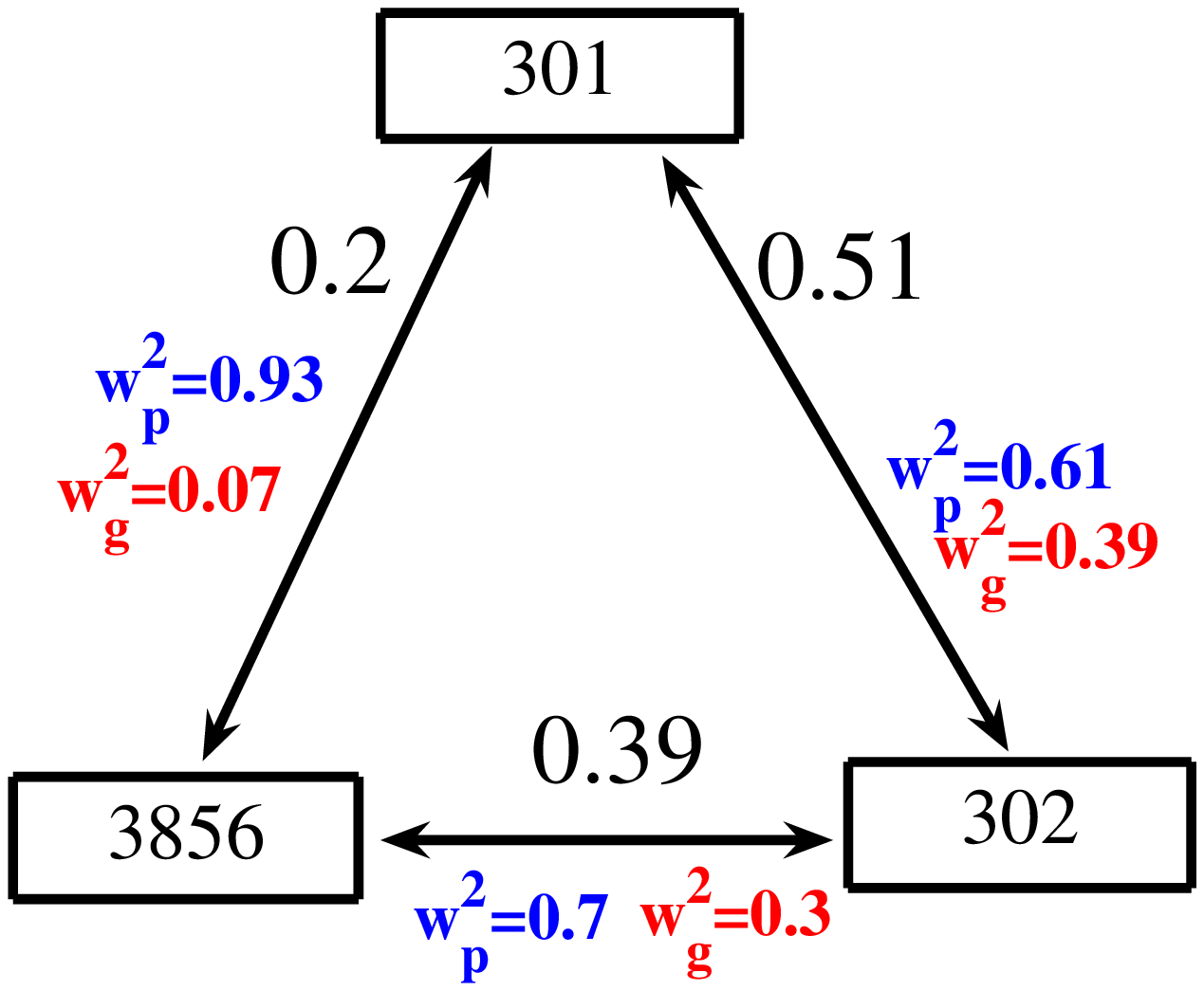}
 \label{fig:toynetworks_rho}
 \end{figure}

Before discussing the full networks we obtained, consider the small illustrative example introduced in Section~\ref{sec:motivation}, involving the three proteins (Annexin A1, Annexin A2, and Keratin 8) and their three corresponding genes (ANXA1, ANXA2, and KRT8).  Figure \ref{fig:toynetworks_rho} shows these subnetworks, now annotated with the values of their estimated correlations and, in the case of the gene/protein network, the canonical weights as well.  As one can easily observe, the protein and gene networks differ in the values of their (marginal) correlations and, consequently,  in their structure. For example, the correlation between proteins Annexin 1 and Keratin 8 is negative, $-0.18$, but, nevertheless, sufficient to produce an edge in the network; the correlation between the corresponding genes ANXA1 and KRT8 is positive, $0.03$, but insufficient to declare an edge. At the same time, the absolute value of the canonical correlation, based on the combined expression profiles, is equal to $0.2$.  Furthermore, examining the canonical weights on this edge, we see that $93\%$ of the canonical correlation can be explained by protein-level information, while only $7\%$ is explained by gene-level information.

 This example is suggestive in two ways. First is that different molecular profiles can produce different networks; and second is that the network inferred from combined molecular profiles via canonical correlation can effectively summarize the combined contributions of the two types of measurements. 

Now consider the networks comprised of the full set of 91 nodes.  Table~\ref{tab:bionet_summaries} reports the number of edges declared  for each network, and the corresponding network densities, while Table~\ref{tab:jsimilarities} summarizes the extent to which edges are shared between networks, through both the Jaccard similarities and the raw counts.  We see that the gene-protein network has the largest number of edges (791), with a density of almost $0.20$, while the protein and gene networks have noticeably fewer edges (426 and 240, respectively), with densities roughly half and a quarter that of the gene-protein network.  Furthermore, the gene-protein network shares over 40\%  of its edges (329) with the protein network, but only about 25\% with the gene network.  In contrast, the protein and gene networks themselves share comparatively few edges (52).  Most interestingly, the gene-protein network contains $309$ edges that are unique and belong to neither the protein nor the gene networks. The presence of such edges indicates both high correlation of between gene and protein profiles for the same node and/or high cross-correlation of gene and protein profiles for distinct nodes. 
\begin{table}[h!]
\centering
 \begin{tabular}{c|r|r|r}
 ~ & { protein network} & {gene network} & {gene-protein network} \\ 
\hline
Nodes($N_v$) & 91 & 91 & 91 \\
\hline 
Edges($N_e$) & 426&   240 &   791 \\
\hline
Density & 0.10 & 0.06 & 0.19 \\
\hline
LCC & 90 & 80 & 91 \\
\hline
Avg Correlation($\hat{\rho}$) & 0.26 &    0.18 &   0.53 \\
\hline
Avg Degree ($\bar{d} $) & 9.36 &  5.27 & 17.38\\
\hline
Avg Clustering & 0.36 &    0.31 &   0.39\\
\hline
Avg Betweenness & 0.034 &   0.041 &   0.022\\
\hline
~ & ~ & ~ & ~ \\
 \end{tabular}
\caption{Summary statistics for protein, gene, and gene-protein networks: number of nodes, number of edges, density, size of the largest connected component (LCC), average nonzero correlation, degree, clustering coefficient, and (normalized) betweenness centrality.}
\label{tab:bionet_summaries}
\end{table}

\begin{table}[h!]
\centering
 \begin{tabular}{c|r|r|r}
 %\multicolumn{2}{|c|}{tbd}\\
 %\hline
 ~ & {protein network} & {gene network} & {gene-protein network} \\ 
\hline
 {Protein Network} &  1.0 (426) & 0.09(52) & 0.37(329) \\
\hline 
{Gene Network}  & ~ & 1.0 (240)  &  0.25(205) \\
\hline
{Gene-Protein Network} & ~ & ~ & 1.0 (791)  \\
\hline
 ~ & ~ & ~ & ~\\
 \end{tabular}
\caption{Jaccard similarities (number of shared edges) between gene, protein, and gene-protein networks.}
\label{tab:jsimilarities}
\end{table}         

 Also shown in Table~\ref{tab:bionet_summaries} are other standard summaries of network structure, including the size of the largest connected component and the average degree, clustering coefficient, and betweenness centrality. 
We refer the reader to~\citet[Chap. 4]{Kolaczyk} for definitions.  We see that only the gene-protein network is fully connected.  In addition, the average degree of nodes in the gene-protein network is nearly twice that in the protein network and over three times that in the gene network.  Furthermore, while the protein and gene-protein networks display similar levels of clustering (i.e., proportions of triads closing to form triangles), the gene network shows somewhat less.  On the other hand, all three networks show similar levels of betweenness centrality.  Particularly interesting, however, is the fact that the gene-protein network shows some evidence for a bimodal degree distribution, suggesting that there are potentially two classes of nodes in the network.  Note that the spikes at zero in the histograms of degree, clustering, and betweenness for the gene network are due to isolated nodes.
 \begin{figure}[!h]
 \centering
 \caption{Distribution of degree (top row), clustering coefficient (middle row), and (normalized) betweenness centrality (bottom row), for the protein (left column), gene (middle column), and gene-protein (right column) networks.}
 \includegraphics[width=1.6in, height=1.2in]{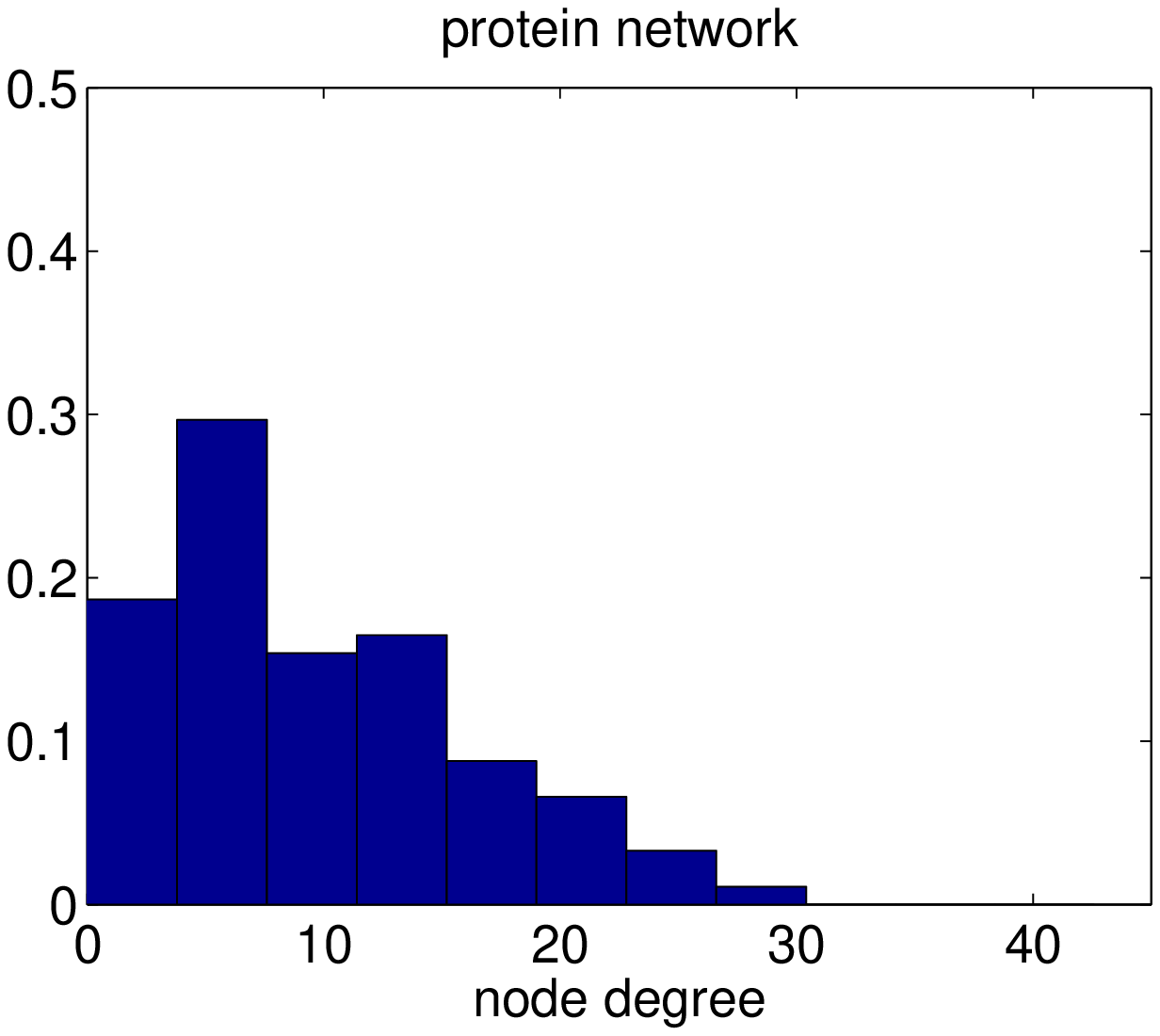} \hskip 0.01in
\includegraphics[width=1.6in, height=1.2in]{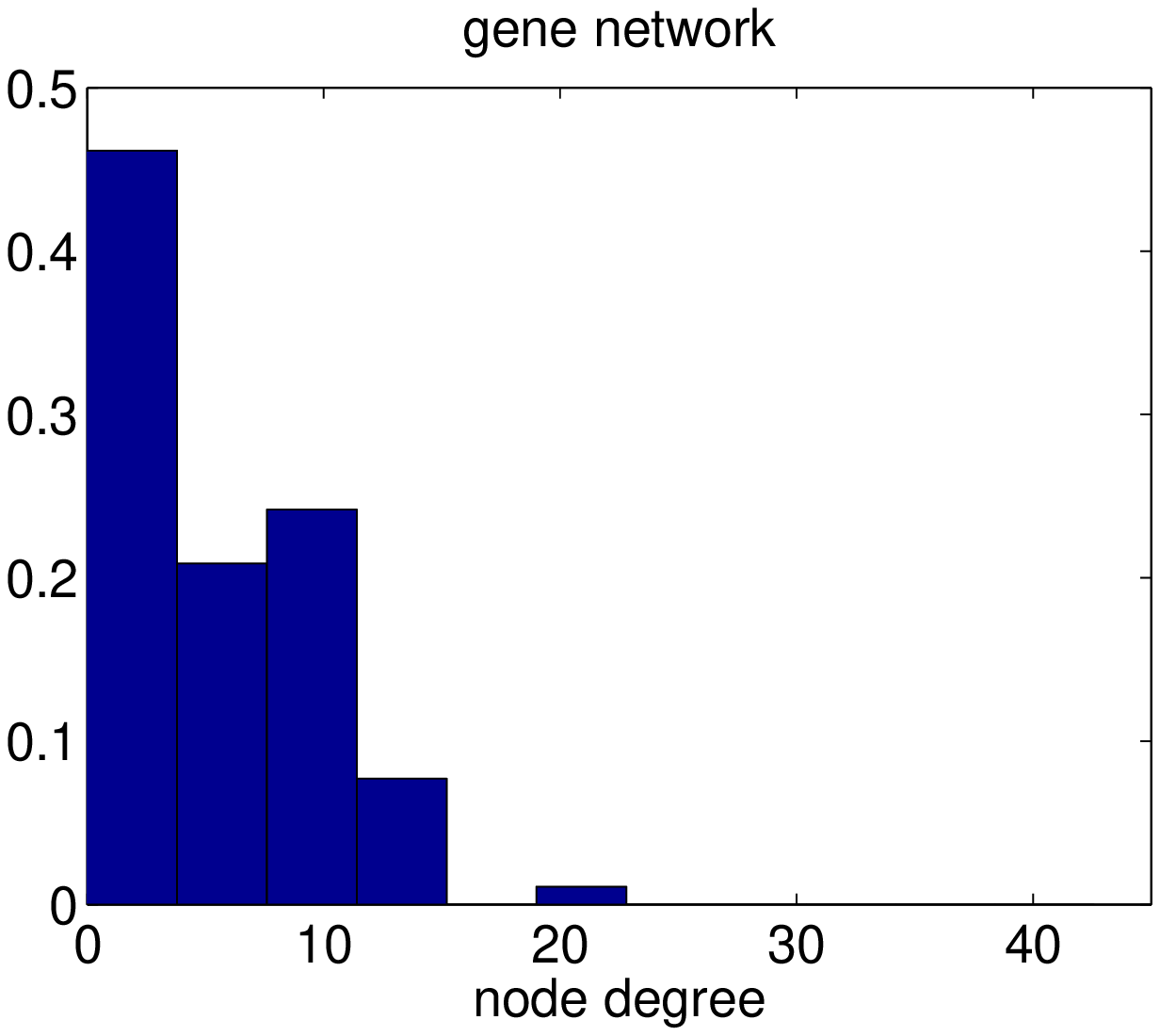} \hskip 0.01in
 \includegraphics[width=1.6in, height=1.2in]{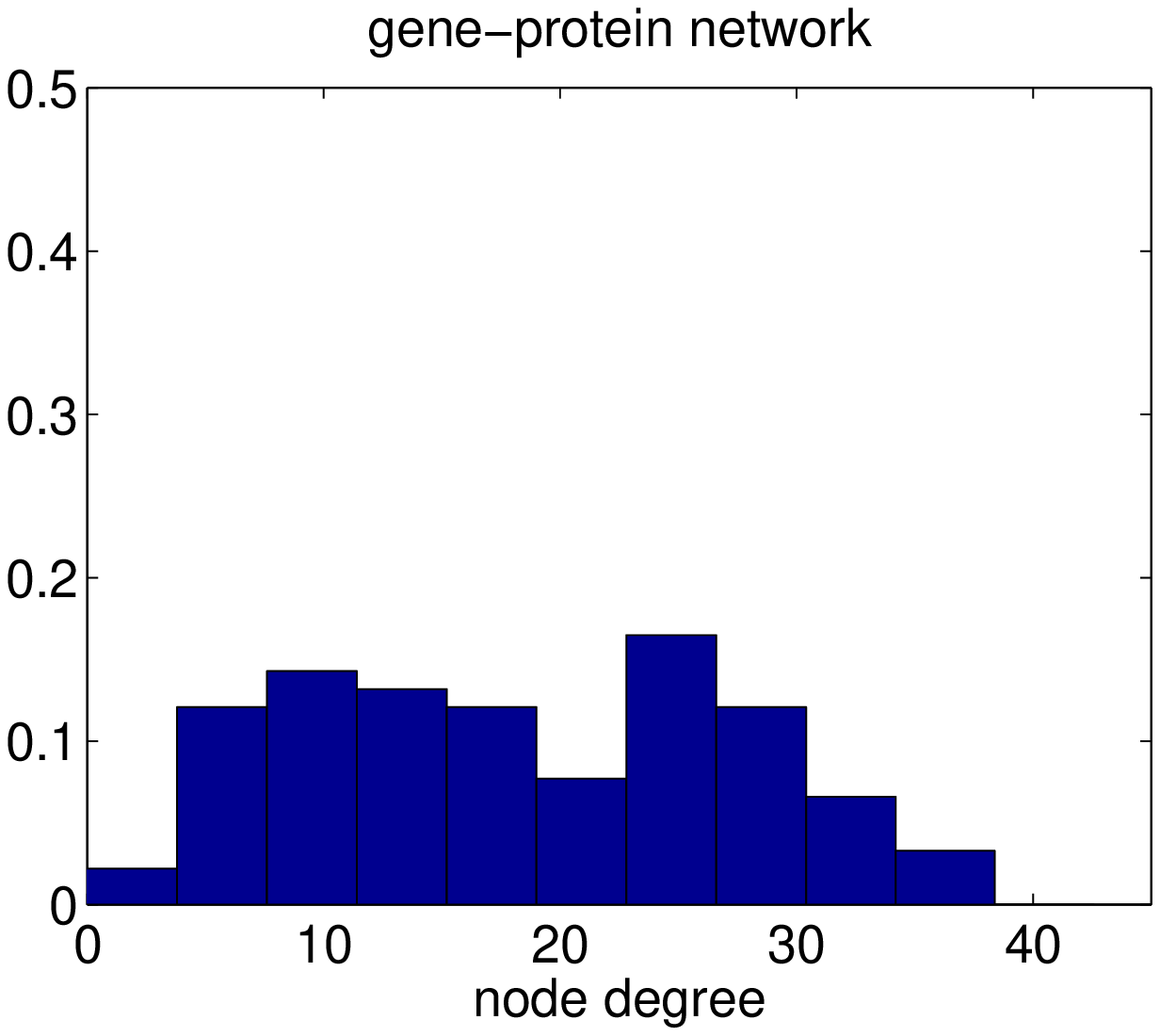} \hskip 0.01in \\
 \includegraphics[width=1.6in, height=1.2in]{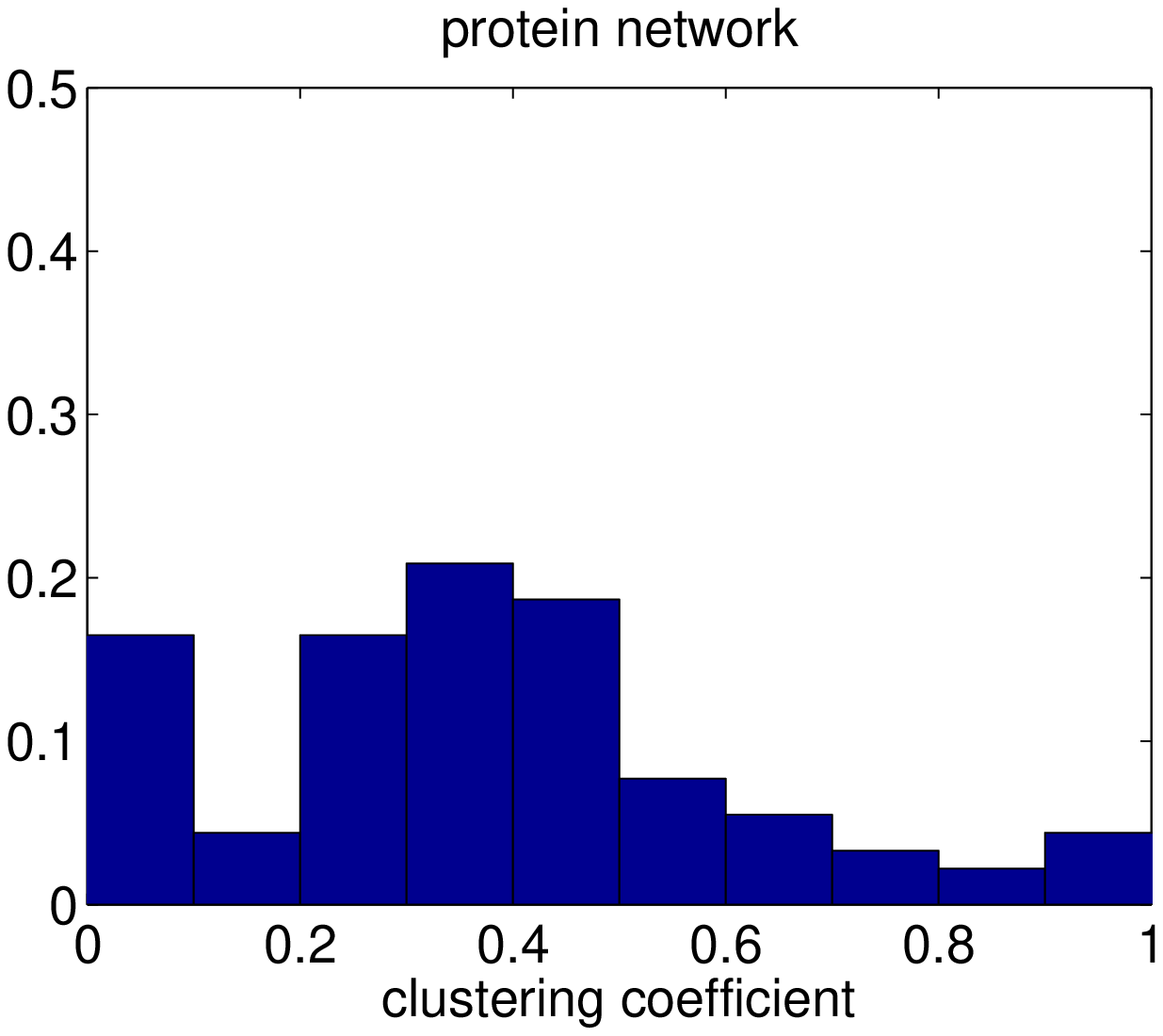} \hskip 0.01in
\includegraphics[width=1.6in, height=1.2in]{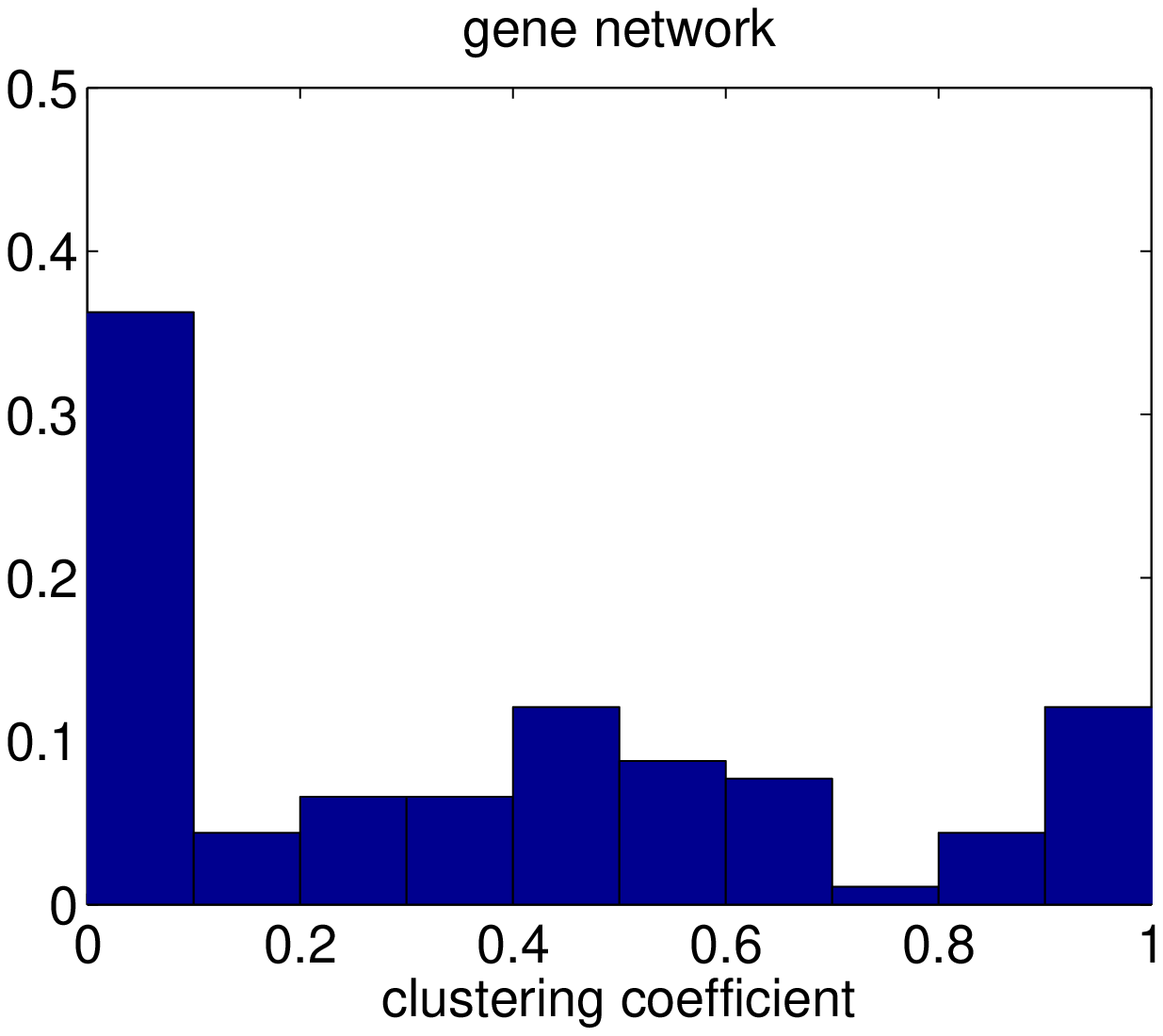} \hskip 0.01in
 \includegraphics[width=1.6in, height=1.2in]{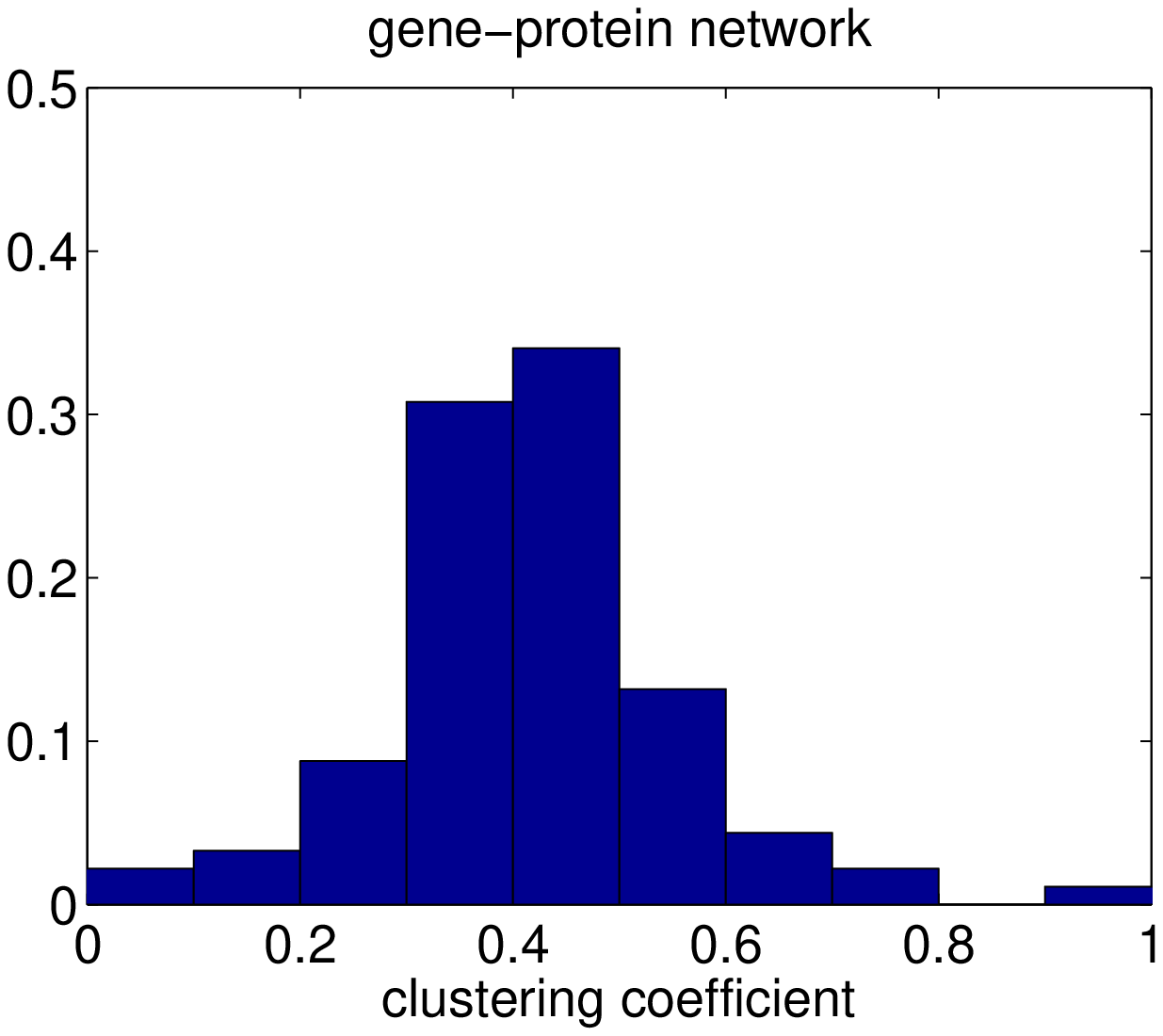} \hskip 0.01in \\
 \includegraphics[width=1.6in, height=1.2in]{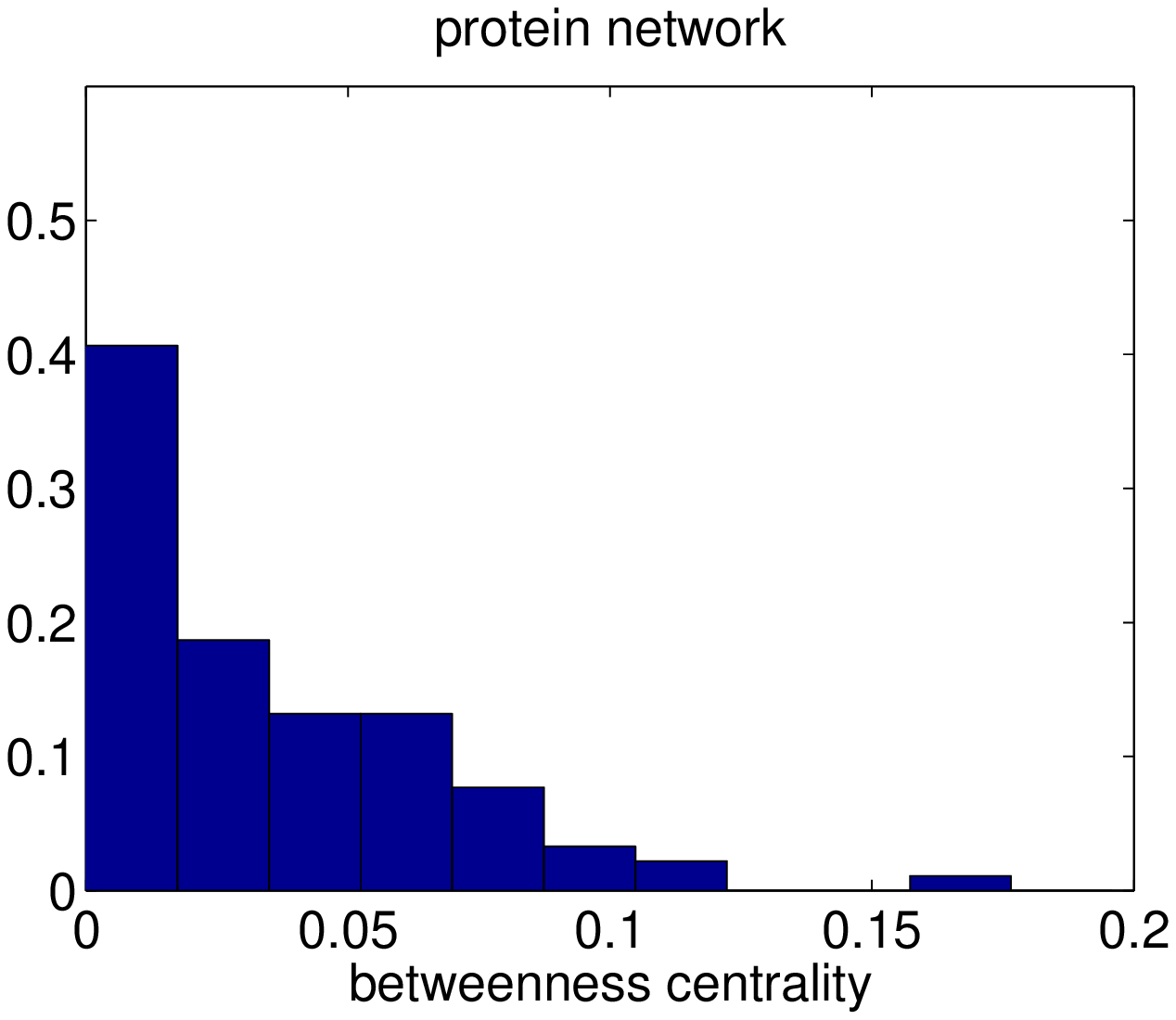} \hskip 0.01in
\includegraphics[width=1.6in, height=1.2in]{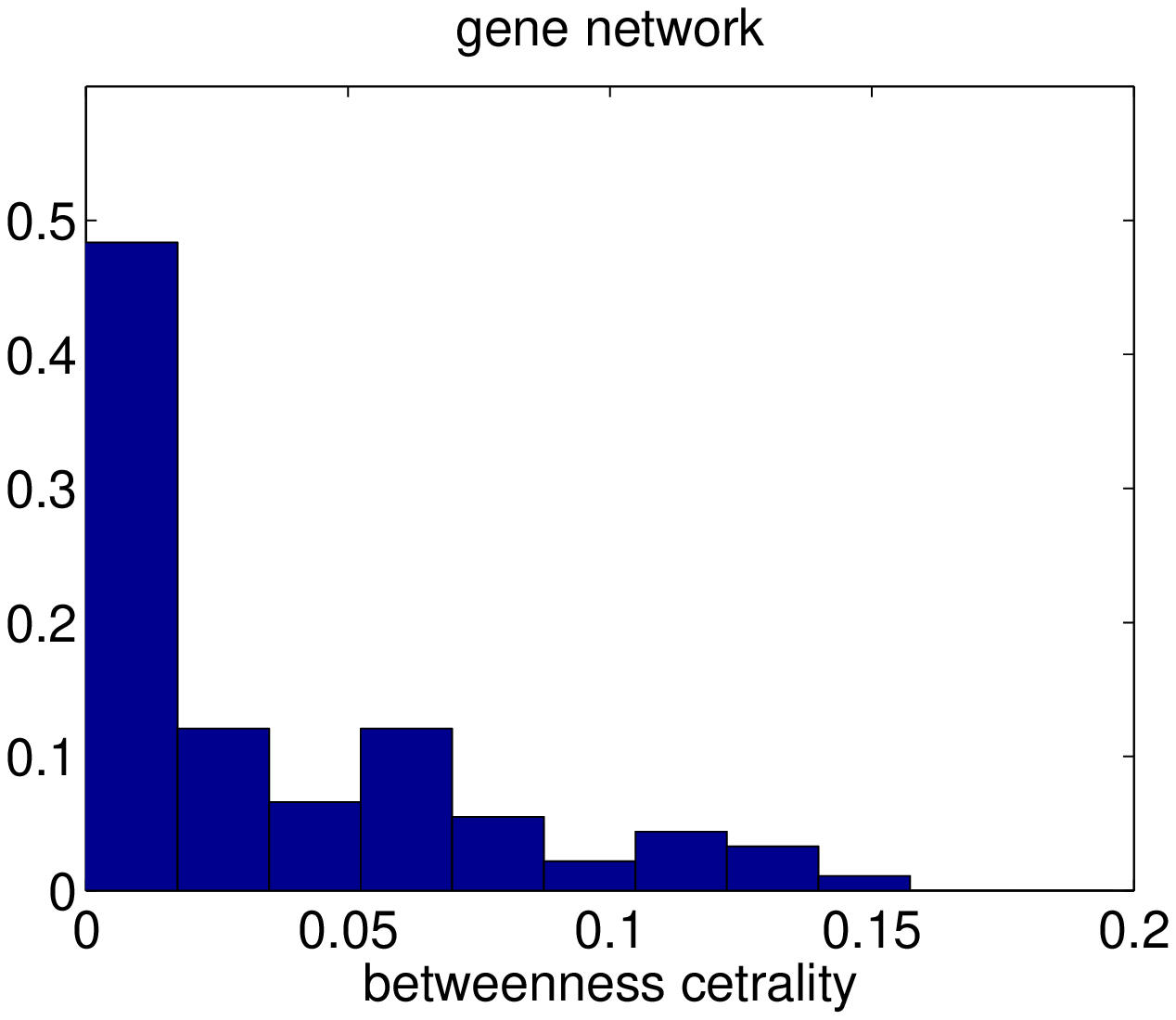} \hskip 0.01in
 \includegraphics[width=1.6in, height=1.2in]{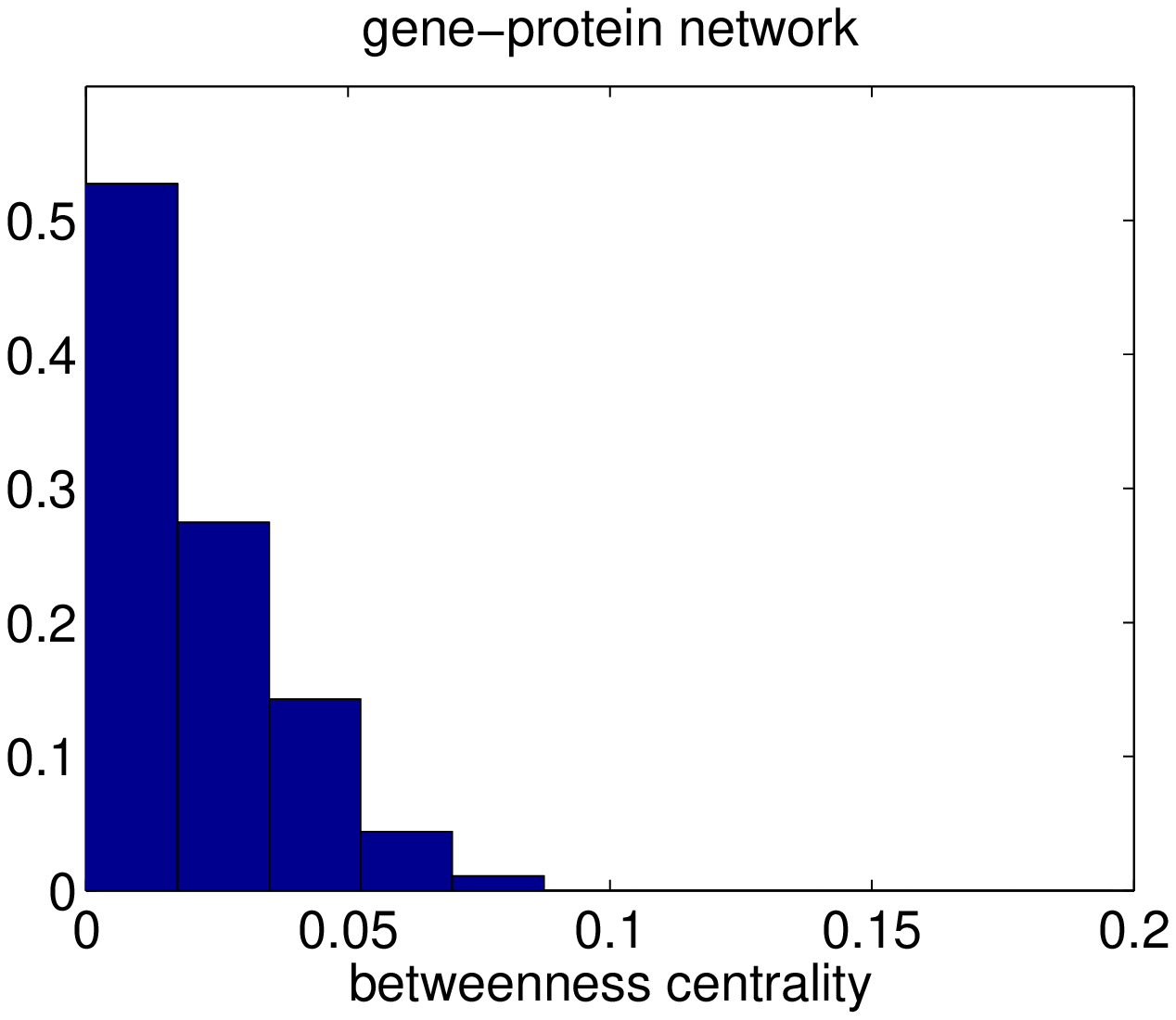} \hskip 0.01in
 \label{fig:bionets_degree}
 \end{figure} 

\subsection{Edge and Node Classification}
\label{sec:classification}

We now focus on analysis of the gene-protein network alone, with the specific goal of better understanding the contribution of the two node attributes (i.e.,  gene expression and protein profile) to the edges incident to each node.  We separate edges/nodes into three separate classes using a simple classification heuristic based on the canonical weights.  Alternatively, we also tried using more sophisticated methods of 'community detection' but found that the results obtained were substantially less interpretable.

\begin{figure}[!htbp]
 \centering
 \caption{Distribution of the canonical weights (squared) corresponding to gene-protein network.}
 \includegraphics[width=4.2in, height=2.4in]{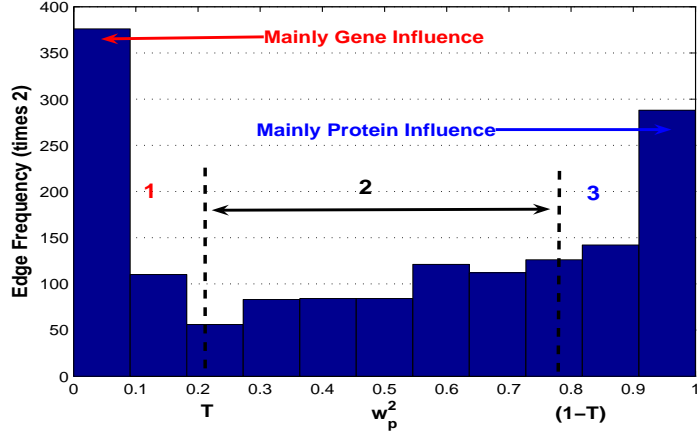}
 \label{fig:bionets_weights}
 \end{figure}

In our analysis, for each pair of nodes with a declared edges, we take the vector of canonical weights, say $w_p$ and $w_g$, corresponding to protein and gene attributes, respectively, and standardized them to have unit length.  A plot of the values $w_p^2$, over all edges, is shown in Figure~\ref{fig:bionets_weights}.  The distribution shows two clear peaks at the far left and right extremes, corresponding to $w_p^2$ close to zero and one, respectively.  The remainder of the distribution between the two peaks is relatively flat.  These observations suggest separating edges into three classes, through the use of a threshold, say $T\in (0,1)$, with edges for which $0 \le w^2_p \le T$ described as mainly gene-influenced, edges for which $1-T\le w^2_p \le 1$, as mainly protein-influenced, and the rest as being of mixed type.  By extension, we then similarly classify nodes according to the majority class of its incident edges.

\begin{figure}[h!]
\centering
\includegraphics[width=1.5in, height=1.0in]{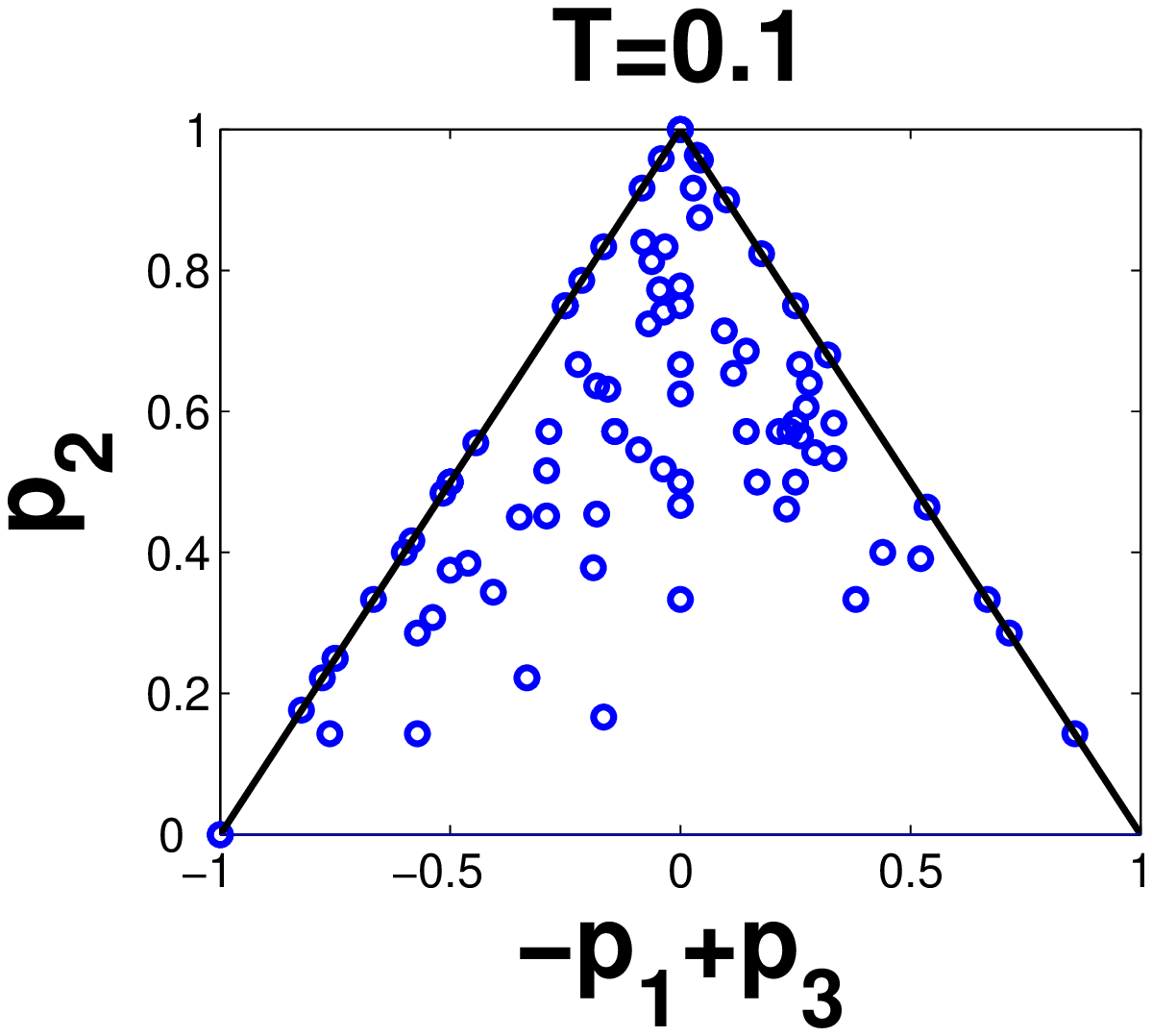}
\includegraphics[width=1.5in, height=1.0in]{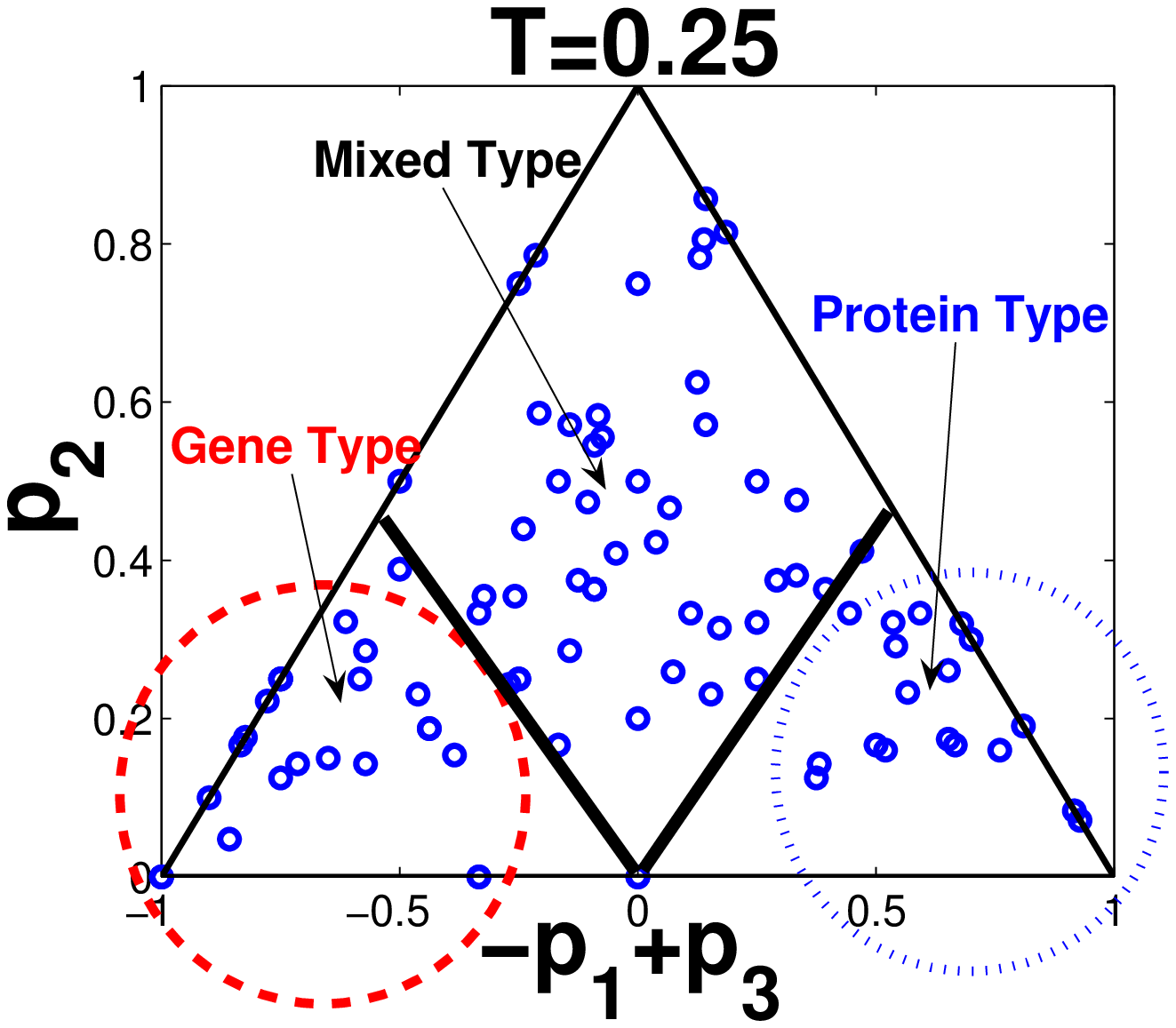}
\includegraphics[width=1.5in, height=1.0in]{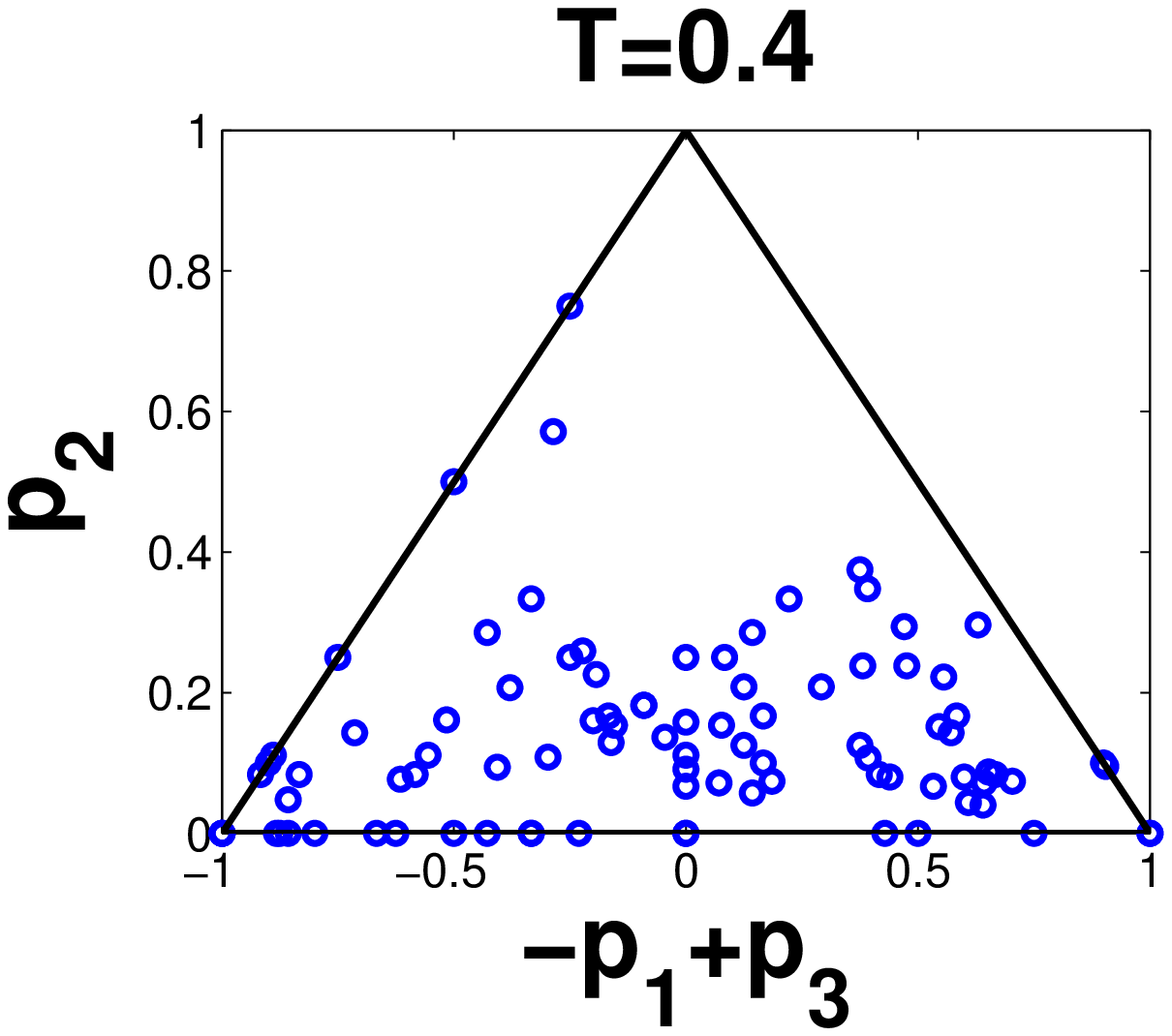}
\caption{Node classification, according to proportion of gene / protein influence on incident edges.}
\label{fig:node_classes}
\end{figure}

Figure~\ref{fig:node_classes} provides a visual illustration of the same process of node classification, for the choices of threshold $T=0.1, 0.25$, and $0.4$.  For each node the proportions $p_{gene}$, $p_{protein}$, and $p_{mixed}$ incident edges were computed. Because the sum of these proportions is one, the nodes may be conveniently displayed in the unit simplex.  Nodes that are close to the bottom left corner have a large proportion of gene edges, while those that are close to the bottom right corner have a large proportion of protein edges.  Mixed nodes tend to be located near the top corner.  Therefore, the location of each node is an indication of the contribution of each of the two attributes to its connectivity in the gene-protein network.  Based on visual inspection of Figures~\ref{fig:bionets_weights} and~\ref{fig:node_classes}, we chose a threshold of $T=0.25$ as most reasonable and use that in the remainder of our analysis, described below.

Note that the above-described approach for classifying nodes can be extended in a natural manner when there are $K>2$ attributes.  First, one separates edges/nodes into $K+1$ separate classes using the canonical weights. Specifically, for each pair of nodes with a declared edge, the vector of canonical weights $w_1,\,w_2, \ldots, w_K$, corresponding to each of the $K$ attributes, are standardized and the maximum of the corresponding squared values is noted, say $w_l^2$. Through the use of a threshold $T\in (0,1)$, an edge is characterized as mainly influenced by this attribute $l$ if $1-T\le w_l^2 \le 1$; otherwise the edge is characterized as being of mixed type. A node can then be classified according to the majority class of its incident edges via the use of a multidimensional analogue of our triangular strategy.  In particular, for each node, proportions $\{p_{attr_l}\}$ and $p_{mixed}$ need to be computed and then analyzed on the multidimensional unit simplex. Note that nodes in 'bottom' corners will correspond to groups of nodes mostly effected by a single attribute, while all mixed-type nodes will be concentrated near the 'top' corner.

\subsection{Biological Interpretation}
\label{sec:enrichment}

Our classification analysis provides an ability to suggest a primary 'role' in which each node participates in the biology underlying our measurements, that is, either at the level of gene expression, protein expression, or both.  In order to assess the extent to which such assignments may be biologically meaningful, we perform an enrichment analysis of our three classes of genes/proteins against the biochemical pathways in the Kyoto Encyclopedia of Genes and Genomes (KEGG) pathways \citep{Kanehisa_2002}.  That is, we identify those cases in which our classes contain significant overlap with particular collections of genes related by their common participation in various specific biochemical processes and, through our understanding of those processes, offer an interpretation of the assignments produced by our classification.

A preliminary comparison of our $91$ network nodes with KEGG revealed that only $68$ of the corresponding genes were contained in at least one of the $148$  KEGG pathways.  More specifically, $15$ protein nodes,  $18$ gene nodes, and $35$ mixed nodes were represented in KEGG.  See the Appendix, Table~\ref{tab:node_classification}.  Accordingly, our enrichment analysis is restricted to this subset of nodes.  For each pathway and each class, we performed a standard hyper-geometric test (i.e., a so-called test for enrichment in the computational biology literature) of independence for allocation of the genes in that class between the, say, $M$ genes in the pathway and the remaining $5017-M$ KEGG genes outside that pathway.  A class is said to be 'enriched' for a given pathway if the null hypothesis is rejected.  To adjust for multiplicity due to the  large number of KEGG pathways, we again use the \cite{Benjamini_1995} false discovery rate (FDR) control procedure and set $\gamma=0.05$. Note that prior to conducting our  tests, we excluded from the analysis all KEGG pathways related to any type of cancer or any other disease, in general, restricting our focus to only those pathways involved with more specific biological functions.

In examining our results, we find that the protein nodes are enriched for 14 pathways, the gene nodes are enriched for one pathway, and the mixed nodes are enriched for 37 pathways.  See the Appendix, Table~\ref{tab:enrichment}.  The pathways for which the protein nodes are enriched are almost all involved with signaling activity (e.g., JAK-STAT-SIGNALING, INSULIN-SIGNALING, GNRH-SIGNALING),  for which we can expect to see coordinated activity at the level of protein expression.  The pathway for which the gene nodes are enriched is called MISMATCH REPAIR, which refers to the process whereby mismatches that may occur during DNA replication and recombination are repaired.  This pathway also is among the 14 pathways enriched by our protein nodes.  However, it makes sense that we would see enrichment as well with nodes associated primarily at the level of gene expression, due to the intimate connection between replication and gene transcription/translation. Finally, we note that the set of nodes classified as being of mixed status are enriched for 24 KEGG pathways.  These include MISMATCH REPAIR and 12 of the other pathways with which the protein nodes were enriched, but also include, for example, various metabolic pathways (e.g., RIBOFLAVIN-METABOLISM), thus seeming to confirm the appropriateness of the label 'mixed'. 

\section{Concluding Remarks}
\label{sec:remarks}  
  
 In this paper, we proposed to use canonical correlation to incorporate multiple node attributes and measure a total similarity between nodes pairs in association networks. Using estimated canonical weights, we assessed the importance of individual node attributes and examined both analytically and numerically the impact of partial information (i.e., measurements of only some, but not all, attributes) on the ability to detect an edge between two nodes.   More generally, we also examined the impact of attribute selection on higher-level network summary statistics, such as degree distribution, and betweenness centrality. For the special case of a network with two attributes collected for each node, we proposed a simple heuristic to characterize network edges and group nodes with respect to the influence of each attribute. We evaluated the proposed framework in the context of gene/protein regulatory networks in human cancer cells, and found that a network based on combined protein profiles and gene expressions appears to be a considerably more rich summary of information than one defined on only a single molecular profile alone.   

Our work was developed with an assumption of continuous measurements.  While in principle it is true that often categorical measurements can be transformed to the continuous case in a useful manner, a more satisfying solution would be an extension of our work based on log-linear models.  Previous work on modeling multiple sociometric relations (e.g., \cite{Fienberg_1985}) should be instructive here.

As noted earlier, topology inference in association networks typically is done using either hypothesis testing or regression methods~\cite[Chap. 7.3]{Kolaczyk}.  A regression-based analogue of the work presented here would be welcome.  Such an approach would presumably exploit the connection between canonical correlation and multiple regression.  But given the large number of variables entering such a regression (e.g., one for each node being considered as a neighbor for a fixed node of interest), some appropriate form of penalization will be critical.  

Last, we mention that while we focused here largely on the case of just two node attributes, the other extreme, in which the number of attributes $K$ is very large, is also likely to be of considerable interest.  In particular, there are likely interesting connections between this case and the current body of work on high-dimensional inference and sparseness, given that in reality a large set of $K$ measured attributes does not necessarily mean that any more than a few are actually important drivers of association between nodes.

 \bibliographystyle{imsart-nameyear}
 \bibliography{biblio}

\section{Appendices}
\label{sec:appendix}

\subsection{Biological Interpretation Tables}
Our classification analysis provides an ability to suggest a primary 'role' in which each node participates in the biology underlying our measurements. 
\begin{table}[b!]
\centering
\label{tab:node_classification}
 \begin{tabular}{p{0.6in}|p{0.8in}|p{1.0in}|p{2.0in}}
Nodes & {Protein Type} & {Gene Type} & {Mixed Type} \\ 
 \hline
%\multicolumn{2}{|c|}{Nodes}\\
%\hline
% N & 20 & 28 & 43\\ \hline
% N in KEGG & 15 &  18 & 35 \\ \hline
Contained in KEGG & CDH1, CDK4, CDK5, CDK7, FN1, GRB2, MSH6, GTF2B, HRAS, IRS1, JAK1, STAT1, STAT6, IRF9, RNASEH2A &

ACVR2A, FASLG, CDH3, CDK6, ERBB2, MCM7, CD46, MLH1, MSH2,  MSN, NCAM1, PRKCH, PRKCI, MAP2K2, TGFB1I1, VASP, RIPK1, EXOC4 &

PARP1, CASP7,  CCNA2,  CCNB1,  CDH2,  CDKN2A,  AP2M1,  CRK, CTNNB1,  CTTN,  EP300,  XRCC6,   GSK3B,  GSTP1,  HSPA4,   HSPD1, NME1,  PCNA,  PGR, PRKCA, PRKCB, MAPK1, MAP2K1,PTPN6, PTPN11, RB1, RELA, STAT3, STAT5A, TP53, TUBB2A, TYR, EZR,  RADD, FADD
 \\
\hline
NOT contained in KEGG & ANXA4,  CDC2, KRT8, MGMT, ADNP &
ANXA1, ANXA2, KLK3, CASP2, DSG1, ESR1, KRT7, KRT19, AKAP5,  AKAP8 &
KRT18, MCC,   PRSS8, ATXN2, SMARCB1, VIL1,  MVP, KRT20 \\
\hline
~ & ~ & ~\\
 \end{tabular}
\caption{Preliminary comparison of $91$ network nodes with KEGG revealed only $68$ contained in at least one of the $148$  KEGG pathways: $15$ protein nodes,  $18$ gene nodes, and $35$ mixed nodes.}

\end{table}
\begin{table}[t!]
\centering
 \begin{tabular}{l|c|c|c}
 KEGG Pathway & Gene Type & Protein Type & Mixed Type \\ 
 \hline
MISMATCH-REPAIR & X & X & X \\
JAK-STAT-SIGNALING-PATHWAY & ~ & X & X\\
T-CELL-RECEPTOR-SIGNALING-PATHWAY & ~ & X & X\\
NEUROTROPHIN-SIGNALING-PATHWAY & ~ & X & X\\
INSULIN-SIGNALING-PATHWAY  & ~ & X & X\\      B-CELL-RECEPTOR-SIGNALING-PATHWAY & ~ & X & X\\
FC-EPSILON-RI-SIGNALING-PATHWAY & ~ & X & X\\
CHEMOKINE-SIGNALING-PATHWAY & ~ & X & X\\
ERBB-SIGNALING-PATHWAY & ~ & X & X\\
GAP-JUNCTION & ~ & X & X\\
DORSO-VENTRAL-AXIS-FORMATION & ~ & X & X\\
FOCAL-ADHESION & ~ & X & X\\
GNRH-SIGNALING-PATHWAY & ~ & X & X\\
DNA-REPLICATION  & ~ & X & ~ \\ 
TIGHT-JUNCTION & ~ & ~ & X\\
MELANOGENESIS & ~ & ~ & X\\
CELL-CYCLE & ~ & ~ & X\\
LONG-TERM-POTENTIATION & ~ & ~ & X\\
PROGESTERONE-MEDIATED-OOCYTE-MATURATION & ~ & ~ & X\\
APOPTOSIS & ~ & ~ & X\\
NATURAL-KILLER-CELL-MEDIATED-CYTOTOXICITY & ~ & ~ & X\\
FC-GAMMA-R-MEDIATED-PHAGOCYTOSIS & ~ & ~ & X\\
WNT-SIGNALING-PATHWAY & ~ & ~ & X\\
ADIPOCYTOKINE-SIGNALING-PATHWAY  & ~ & ~ & X\\
LEUKOCYTE-TRANSENDOTHELIAL-MIGRATION & ~ & ~ & X\\
ADHERENS-JUNCTION & ~ & ~ & X\\
 VEGF-SIGNALING-PATHWAY & ~ & ~ & X\\
ALDOSTERONE-REGULATED-SODIUM-REABSORPTION & ~ & ~ & X\\
MAPK-SIGNALING-PATHWAY & ~ & ~ & X\\
TOLL-LIKE-RECEPTOR-SIGNALING-PATHWAY & ~ & ~ & X\\
OOCYTE-MEIOSIS & ~ & ~ & X\\ 
VASCULAR-SMOOTH-MUSCLE-CONTRACTION & ~ & ~ & X\\ 
P53-SIGNALING-PATHWAY & ~ & ~ & X\\
RIG-I-LIKE-RECEPTOR-SIGNALING-PATHWAY & ~ & ~ & X\\ 
BASE-EXCISION-REPAIR & ~ & ~ & X\\
NON-HOMOLOGOUS-END-JOINING & ~ & ~ & X\\
RIBOFLAVIN-METABOLISM & ~ & ~ & X\\
NOD-LIKE-RECEPTOR-SIGNALING-PATHWAY  & ~ & ~ & X\\
\hline
~ & ~ & ~ & ~\\
 \end{tabular}
\caption{Results of enrichment analysis: protein type nodes are enriched for 14 pathways, the gene nodes - for one pathway, and the mixed nodes - for 37 pathways.}
\label{tab:enrichment}
\end{table}

\subsection{Proposition Proof}
Here we show that if the assumption of equal marginal covariance matrices ($\Sigma_{ii} = \Sigma_{jj} = \Sigma_m$) and symmetrical cross-covariance matrix ($\Sigma_{ij} = \Sigma_{ji} = \Sigma_c$)  for two nodes $i$ and $j$ are satisfied, then optimization problem \eqref{eqn:cancorr_opt} can be simplified to:
\begin{equation}
 \rho_c(i,j) =\max_{w\in \mathbb{R}^{|C|}}\frac{w^{T}\Sigma_c w}{w^{T}\Sigma_m w}, 
\end{equation}
and only one set of weights for each edge $e(i,j)$ is required. To proof that, we first observe that solution to the problem is not affected by rescaling $w_i$ or $w_j$ either independently or together, that is, if replacing $w_i$ by $\alpha w_i$ and $w_j$ by $\beta w_j$, canonical correlation $\rho(i,j)$ would not change:
\begin{eqnarray}
 \rho_c(i,j) &=& \max_{w_i,w_j \in }\frac{\alpha w_i^{T}\Sigma_c \beta w_j}{\sqrt{\alpha w_i^{T}\Sigma_m \alpha w_i}\sqrt{\beta w_j^{T}\Sigma_m\beta w_j}} \nonumber\\
&=& \max_{w_i,w_j}\frac{w_i^{T}\Sigma_c w_j}{\sqrt{w_i^{T}\Sigma_m w_i}\sqrt{w_j^{T}\Sigma_m w_j}}, \mbox{ for all } \alpha, \beta \in R. \nonumber
\end{eqnarray}   
 Therefore, the canonical optimization problem \eqref{eqn:cancorr_opt} is equivalent to:
\begin{eqnarray}
\max_{w_i,w_j} w_i^{T}\Sigma_c w_j, \mbox{  subject to}\\
w_i^{T}\Sigma_m w_i = 1,~~ w_j'\Sigma_m w_j = 1. \nonumber
\label{eqn:constraints}
\end{eqnarray}   
 Applying the method of Lagrange multipliers, we construct a maximization criterion as
$$
L(\lambda_i, \lambda_j, w_i, w_j) =  w_i^{T}\Sigma_c w_j - \frac{\lambda_i}{2}(w_i^{T}\Sigma_m w_i -1) - \frac{\lambda_j}{2}(w_j^{T}\Sigma_m w_j -1).
$$
 Taking partial derivatives of $L(\lambda_i, \lambda_j, w_i, w_j)$ with respect to $w_i$ and $w_j$, we obtain the following system of equations \eqref{eqn:lagrange_system}:
\begin{eqnarray*}
\Sigma_c(i,j) w_j - \lambda_i \Sigma_m(i) w_i &=& 0,  \\ 
\Sigma_c^{T}(i,j) w_i - \lambda_j \Sigma_m(j) w_j &=& 0.
\end{eqnarray*}   
 Multiplying the first equation by $w_i^{T}$ and the second equation by $-w_j^{T}$ and adding them together, we have
$$-\lambda_i w_i^{T}\Sigma_m w_i + \lambda_j w_j^{T}\Sigma_m w_j =0, $$    which together with constraints implies $\lambda_i=\lambda_j=\lambda$. In this case, we may reduce the system \eqref{eqn:lagrange_system} to the system
\begin{eqnarray*}
\Sigma_c w_j &=& \lambda_i^2 \Sigma_m (\Sigma_c^{-1})^{T} \Sigma_m w_j \\ \Sigma_c^{T} w_i &=& \lambda_i^2 \Sigma_m \Sigma_c^{-1} \Sigma_m (i)w_i,
\end{eqnarray*}   
or assuming $\Sigma_m=\Sigma_m$ and $\Sigma_c^{T}=\Sigma_c$:
\begin{eqnarray*}
\Sigma_c w_j = \lambda_i^2 \Sigma_m \Sigma_c^{-1}\Sigma_m w_j , \mbox{ and } \Sigma_c w_i = \lambda_i^2 \Sigma_m \Sigma_c^{-1} \Sigma_m w_i .
\end{eqnarray*}   
 The last set of equations shows that $w_i$ and $w_j$ are both the eigenvectors of matrix $\Sigma_m^{-1}\Sigma_c \Sigma_m^{-1} \Sigma_c$, correspond to the same eigenvalue $\lambda^2$,  and both satisfy constraints \eqref{eqn:constraints}, so that implies $w_i \equiv w_j = w$. Thus, eigenvalue problem \eqref{eqn:lagrange_system}  is reduced to: 
$$
\Sigma_m^{-1}\Sigma_c w = \lambda_i w. 
$$

% AOS,AOAS: If there are supplements please fill:
%\begin{supplement}[id=suppA]
%  \sname{Supplement A}
%  \stitle{Title}
%  \slink[url]{http://lib.stat.cmu.edu/aoas/???/???}
%  \sdescription{Some text}
%\end{supplement}

\end{document}